\documentclass[submit]{epsv8}

\usepackage{graphicx}
\usepackage{lineno}
\usepackage{textgreek}
\usepackage{amsmath}
\nolinenumbers
\usepackage{url}
\usepackage{multirow}

\title{Prospects for Mercury Observations with the BepiColombo Laser Altimeter (BELA): Implications for Estimating Surface Roughness from Laser Altimetry}
\author{
  Gaku Nishiyama, Institute of Space Research, Deutsches Zentrum für Luft- und Raumfahrt, Rutherfordstr. 2, 12489 Berlin, Germany / Department of Cosmosciences, Graduate School of Science, Hokkaido University, Kita-10, Nishi-8, Kita-ku, Sapporo 060-0810, Japan, gaku.nishiyama@dlr.de
  \\
  Alexander Stark, Institute of Space Research, Deutsches Zentrum für Luft- und Raumfahrt, Rutherfordstr. 2, 12489 Berlin, Germany, Alexander.Stark@dlr.de
  \\
  Christian Hüttig, Institute of Space Research, Deutsches Zentrum für Luft- und Raumfahrt, Rutherfordstr. 2, 12489 Berlin, Germany, Christian.Huettig@dlr.de
  \\
  Kai Wickhusen, Institute of Space Research, Deutsches Zentrum für Luft- und Raumfahrt, Rutherfordstr. 2, 12489 Berlin, Germany, Kai.Wickhusen@dlr.de
  \\
  Christian Althaus, Institute of Space Research, Deutsches Zentrum für Luft- und Raumfahrt, Rutherfordstr. 2, 12489 Berlin, Germany, Christian.Althaus@dlr.de
  \\
  Thomas Behnke, Institute of Space Research, Deutsches Zentrum für Luft- und Raumfahrt, Rutherfordstr. 2, 12489 Berlin, Germany, Thomas.Behnke@dlr.de
  \\
  Ernst Hauber, Institute of Space Research, Deutsches Zentrum für Luft- und Raumfahrt, Rutherfordstr. 2, 12489 Berlin, Germany, Ernst.Hauber@dlr.de
  \\
  Klaus Gwinner, Institute of Space Research, Deutsches Zentrum für Luft- und Raumfahrt, Rutherfordstr. 2, 12489 Berlin, Germany, Klaus.Gwinner@dlr.de
  \\
  Konrad Willner, Institute of Space Research, Deutsches Zentrum für Luft- und Raumfahrt, Rutherfordstr. 2, 12489 Berlin, Germany, Konrad.Willner@dlr.de
  \\
  Noriyuki Namiki, National Astronomical Observatory of Japan, 2-21-1, Osawa, Mitaka, Tokyo 181-8588, Japan / The Graduate University for Advanced Studies, SOKENDAI, Hayama, Kanagawa, 240-0193, Japan, nori.namiki@nao.ac.jp
  \\
  Frederic Schmidt, Université Paris-Saclay, CNRS, GEOPS, 91405, Orsay, France, frederic.schmidt@universite-paris-saclay.fr
  \\
  Jean Barron, Université Paris-Saclay, CNRS, GEOPS, 91405, Orsay, France, jean.barron@universite-paris-saclay.fr
  \\
  Hugo Lancery, Université Paris-Saclay, CNRS, GEOPS, 91405, Orsay, France, hugo.lancery@universite-paris-saclay.fr
  \\
  Azar Arghavanian, Max Planck Institute for Solar System Research, Justus-von-Liebig-Weg 3, 37077 Göttingen, Germany, arghavanian@mps.mpg.de
  \\
  Hauke Hussmann, Institute of Space Research, Deutsches Zentrum für Luft- und Raumfahrt, Rutherfordstr. 2, 12489 Berlin, Germany, Hauke.Hussmann@dlr.de
}

\abstract{
  Surface reflectance and roughness are key parameters for understanding the geologic evolution of Mercury. The BepiColombo Laser Altimeter (BELA) onboard the ESA–JAXA BepiColombo mission is designed to measure these surface properties, in addition to topography, to provide global information on Mercury’s geology. Accurate prediction of BELA’s performance in surface characterization requires a realistic model of laser return pulse shapes. However, previous performance models have simplified the laser pulse shape with a Gaussian approximation, assuming random roughness on a tilted surface within the footprint. In this study, we develop a comprehensive return-pulse simulation framework for BELA that incorporates non-Gaussian transmitted pulse shapes, analog filtering in the receiver chain, and pulse-shape modification by footprint-scale surface topography. Using high-resolution lunar digital terrain models and synthesized fractal topography as analogs for Mercury, we evaluate BELA performance in range, return pulse energy, and surface roughness measurements. By introducing additional calibration and correction methods, we provide realistic estimates of range and reflectance measurement errors expected in BELA observations starting in 2027. As a result, the range measurements are predicted to have average errors of 1.5 m or less at altitudes below 1000 km. The energy measurement error is estimated to be small enough for distinguishing dark deposits, regolith and exposed ice at altitudes below 900 km. Our simulations further demonstrate that pulse width, which has been widely used as a proxy for footprint-scale roughness, is not a reliable indicator of surface roughness under realistic observing conditions. Instead, we present an alternative approach to constrain surface roughness using digitized return pulse shapes. These results highlight the importance of time-resolved pulse-shape data to constrain surface roughness for BELA and for future planetary laser altimeters.
}


\begin{document}

\maketitle
\nolinenumbers

\section{Introduction}
\label{sec1}
Previous missions exploring Mercury through remote-sensing observations have revealed a wide variety of surface reflectance and compositional characteristics. Reflectance measurements obtained by the Mercury Laser Altimeter (MLA) onboard the Mercury Surface, Space ENvironment, GEochemistry and Ranging (MESSENGER) mission \citep{solomon_messenger_2007,cavanaugh_mercury_2007} identified anomalously dark and bright deposits at 1064 nm inside permanently shadowed regions (PSRs) near the north pole \citep{neumann_bright_2013}. Comparisons with Earth-based radar observations suggest that these anomalies are associated with near-surface water ice, which is either overlain by dark organic-rich materials or directly exposed, depending on the local thermal environment \citep{paige_thermal_2013,neumann_bright_2013,barker_new_2022}. Another distinctive and widespread geomorphological landform on Mercury are the so-called "hollows", defined as a cluster of rimless depressions characterized by a flat floor and steep walls. Images acquired by the Mercury Dual Imaging System (MDIS) onboard MESSENGER \citep{hawkins_mercury_2007} revealed that hollows exhibit high reflectance and a relatively blue spectral signature compared with the surrounding terrain \citep{blewett_multispectral_2009,blewett_hollows_2011}. Moreover, the peripheries of hollows are frequently accompanied by low-reflectance spots composed of volatile-bearing materials \citep{xiao_processing_2021,thomas_hollows_2014}. \\

Combined with reflectance, surface roughness provides important constraints on the formation and evolutionary processes of Mercury’s surface. Using roughness measurements at kilometer-scale baselines derived from MLA along-track data, the Mercurian surface has been broadly classified into two terrain types: cratered terrain and smooth plains \citep{susorney_surface_2017}. This dichotomy reflects the relative contributions of distinct geological processes, such as flood volcanism, impact cratering, and tectonics \citep[e.g.,][]{head_flood_2011,susorney_surface_2018,byrne_mercurys_2014, nishiyama_first_2026,nishiyama_underestimation_2026}. Analyses of MLA-derived digital terrain models (DTMs) with a grid spacing of 125~m further indicate that roughness within polar craters depends on the presence of water ice. Regions exhibiting both dark deposits and radar-bright signatures display significantly lower roughness at baselines of hundreds of meters than neighboring ice-free surfaces, suggesting that surface roughness is reduced by the emplacement of a thick ice layer \citep{deutsch_surface_2022}. In addition, possible variations in roughness around hollows have been inferred from analyses of reflectance ratios between different phase angles using MDIS imagery. Higher ratios observed in dark spots relative to nearby hollow floors and surrounding regolith imply rougher topography at sub-pixel scales; however, these ratios are also influenced by particle size effects \citep{wang_dark_2022}. Disentangling these effects requires quantitative estimates of surface roughness at baselines below several tens of meters, which remain inaccessible due to the limited spatial resolution of MLA coverage and existing image-based DTMs. \\

Global datasets of such surface properties are expected to be acquired by the European Space Agency's (ESA) and the Japan Aerospace Exploration Agency's (JAXA) BepiColombo mission. BepiColombo consists of two spacecraft: the Mercury Planetary Orbiter (MPO) and the Mercury Magnetospheric Orbiter \citep[Mio;][]{benkhoff_bepicolombo_2021,murakami_miofirst_2020}. One of the key instruments onboard MPO is the BepiColombo Laser Altimeter \citep[BELA;][]{thomas_bepicolombo_2019,thomas_bepicolombo_2021}, which will start scientific operations in late March 2027. BELA measures the distance between MPO and the planetary surface by transmitting laser pulses toward Mercury and detecting the returned signals, enabling precise determination of Mercury’s topography, radial tidal deformation, and rotational state. During the nominal one-year mission and mission extensions until 2029, BELA footprints will cover nearly the entire surface of Mercury, with a maximum cross-track spacing of less than 3 km even at the equator \citep{steinbrugge_performance_2018}. These dense range measurements will allow an accurate determination of Mercury's geodetic parameters, providing key constraints on Mercury’s interior structure \citep[e.g.,][]{thor_prospects_2020, desprats_mla_2026}. In addition to range information, BELA records the total energy and temporal shape of each return pulse, which contain information on surface reflectance and footprint-scale topographic properties, such as roughness and local slope. \\

Unlike other previous interplanetary laser altimeters equipped with thresholding detection algorithms, BELA employs a digital detection scheme that finds up to four pulse candidates within a 30-km wide range window. After signal digitization, return pulses are correlated against pre-defined pulse-shape templates characterized by two free parameters. Within each range window, the onboard fitting algorithm identifies four pulse candidates, one of which may correspond to the true surface return when the signal exceeds the noise level. Depending on telemetry allocation, BELA can downlink the best-fit pulse parameters, correlation window (CW) samples composed of 42 signal points equally spaced by 12.5 ns around a pulse candidate, and/or even the samples of the full range window. In Earth orbit, similar pulse shape data have been acquired by the Geoscience Laser Altimeter System (GLAS), where complex return pulses contain information on within-footprint structures such as vegetation \citep{abshire_geoscience_2005,harding_icesat_2005}. Among deep space experiments, BELA is the first laser altimeter to acquire such detailed pulse-shape information from the surface, offering unique opportunities to investigate the reflectance and topography of airless bodies beyond other earlier instruments. \\

Despite the potential scientific value of detailed return-pulse shapes, previous studies have modeled BELA performance assuming simple Gaussian return pulses. As formulated by \citet{gardner_target_1982,gardner_ranging_1992}, the return pulse from a surface governed by random topography can be approximated by a Gaussian shape, similar to that from a flat but sloped surface. Under this assumption, analytical predictions of the BELA system performance have been developed by \citet{gunderson_laser_2006} and \citet{gunderson_bela_2010}. Subsequently, \citet{steinbrugge_performance_2018} refined the return pulse model using semi-analytical simulations with Gaussian pulses synthesized with the noise floor models. Further improvements incorporating nominal spacecraft orbits, updated Mercury topography, and laboratory-measured instrument degradation were introduced by \citet{hosseiniarani_comprehensive_2021}, while still assuming Gaussian return pulses. These models, assuming a  Gaussian pulse shape, may provide a biased performance of BELA because of the assumed simplicity of return pulse shape. For instance, as known for other planetary laser altimeters, the actual return pulses may deviate from simple Gaussian shapes due to the effects of analog bandpass filtering \citep[][see also Section \ref{subsec:filtering}]{smith_lunar_2010} and within-footprint topographies (see also Section \ref{subsec:topography effect}). Ignoring realistic pulse shapes may introduce systematic biases in range, albedo, and roughness measurements of BELA. Therefore, performance assessments as well as calibration methods for BELA need to account for realistic pulse shapes.\\

To refine predictions for BELA’s future performance and to develop methods for characterizing surface roughness and calibrating data, this study simulates return-pulse shapes from realistic Mercurian terrains. By comparing multiple topographic models of Mercury, we first prepare analog topography models of Mercury. We then numerically simulate the modification of laser pulses by surface topography and by electrical processes within the whole receiver chain to model the return-pulse detection by BELA. Next, we evaluate the performance of the pulse-shape characterization using both onboard fitting results and analyses of CW data. Developing a calibration to correct the influence from non-Gaussian pulse shapes, we subsequently assess the impact of return-pulse shape on range and reflectance measurements to update performance predictions. We also demonstrate the feasibility of roughness characterization using pulse width and CW samples, introducing a new method to extract within-footprint roughness information directly from return-pulse shapes. Finally, we discuss implications for future surface characterization by BELA, as well as broader insights into other laser altimetry data.

\section{Topography Model of Mercury}
\label{sec:topography}

\subsection{Analog DTMs}
At present, no DTMs of Mercury with sufficiently high spatial resolution are directly available for our pulse-shape simulations. One of the global DTMs publicly available is the MESSENGER global DTM, which has a grid spacing of 665 m/pix \citep{becker_first_2016}. Using an improved processing scheme developed by \citet{preusker_high-resolution_2017}, higher-resolution DTMs with a grid spacing of 220 m/pix have been generated for individual quadrangles \citep{preusker_toward_2017}. In addition, stereo images acquired by MDIS have been used to produce numerous local- to regional-scale DTMs with a grid spacing down to approximately 50 m \citep{fassett_ames_2016}. However, even these DTMs remain insufficient for our simulations, as their pixel sizes are comparable to or larger than the expected size of BELA footprints. The BELA footprint diameter, defined by the $1/e^{2}$ beam divergence, ranges from approximately 20 to 70 m depending on the altitude of the Mercury Planetary Orbiter (MPO). To quantify pulse-shape modifications induced by within-footprint topography, our simulations require topographic models with spatial resolutions higher than those of any currently available DTMs of Mercury.\\

To compensate for the lack of small-scale topographic information on Mercury, we use a high-resolution lunar DTM as an analog model. Similar to other airless bodies covered with a layer of regolith, both the Moon and Mercury are thought to share surface morphologies governed by similar regolith-related processes. Nevertheless, comparative studies of planetary surface roughness based on laser altimetry have shown that Mercury is smoother than the Moon at baselines longer than several kilometers \citep[e.g.,][]{pommerol_comparative_2012,landais_multifractal_2019,nishiyama_first_2026}. At smaller spatial scales, ranging from meters to sub-kilometers, image-based geological analyses further indicate that Mercury hosts fewer craters and boulders than the Moon \citep{fassett_evidence_2017,kreslavsky_boulders_2021}, and that surface textures like "elephant hide" is absent on Mercury \citep{zharkova_regolith_2020}. These observations have been interpreted as evidence that Mercury possesses a thicker regolith layer than the Moon \citep{kreslavsky_kilometerscale_2014}, implying that Mercury is generally less rough than the lunar surface. Therefore, lunar roughness can be regarded as an upper bound for the global-average roughness of Mercury. In this study, we use a high-resolution DTM of the Apollo-17 landing site from \citet{haase_coordinates_2019} (Figure \ref{fig:roughness_comparison}a; denoted as Apollo-17 DTM hereafter). The Apollo-17 DTM is constructed using images acquired by the Narrow Angle Lunar Reconnaissance Orbiter Camera (LROC) and has a spatial resolution of 1.5 m/pix. This resolution is sufficiently high for our pulse-shape simulations and is adopted as an analog topographic model for Mercury at short baselines. \\

As an alternative approach, we also employ artificially synthesized topography models based on fractals. Since the introduction of fractal concepts into geophysical and geological contexts by \citet{mandelbrot_how_1967}, fractal statistics have been widely used to characterize terrestrial topography. Laser altimetry analyses of other rocky bodies, including the Moon and Mercury, indicate that their surfaces exhibit a power-law scaling behavior over spatial scales ranging from approximately 0.1 to 10 km \citep{landais_universal_2015,landais_multifractal_2019}. Assuming that this fractal property extends to smaller spatial scales, artificial elevation maps can be generated from 2D arrays filled with random numbers, following the procedures described by \citet{schertzer_physical_1987}, \citet{schertzer_non-linear_1991}, and \citet{landais_topography_2019}. By varying the vertical scales of the synthesized topographies, we prepare 11 fractal DTMs covering a range of possible topographies of Mercury (see section \ref{sec: roughness_comparison}). For all those 11 synthetic topographies, we use a Hurst exponent of 0.8 \citep{landais_multifractal_2019} and vary the absolute roughness to cover the range of real Mercury data.\\

\subsection{Roughness Comparison among Analog DTMs}
\label{sec: roughness_comparison}
Quantitative evaluation of surface roughness provides a means of assessing the similarity between the analog DTMs and the actual surface of Mercury. Among the various means proposed for planetary roughness analysis, the root-mean-square (RMS) deviation of elevation differences has been widely used as an indicator that enables quantitative comparison across different geological units \citep[e.g.,][]{shepard_roughness_2001}. On Mercury, RMS deviation measured at kilometer-scale baselines has been used to distinguish cratered terrain from smooth plains \citep[e.g.,][]{susorney_surface_2017}. In this analysis, elevation differences between neighboring two points are computed and binned at each baseline to calculate RMS deviations as a function of baseline. The resulting RMS deviation characterizes the scale dependence of surface roughness and is well suited for describing fractal surfaces, as it follows a power-law relationship with baseline length. In the case of a monofractal surface, the power-law exponent is constant, and the RMS deviation exhibits a linear relationship on a log-log plot. \\

\begin{figure}[ht]
\centering
    \includegraphics[width=0.7\linewidth]{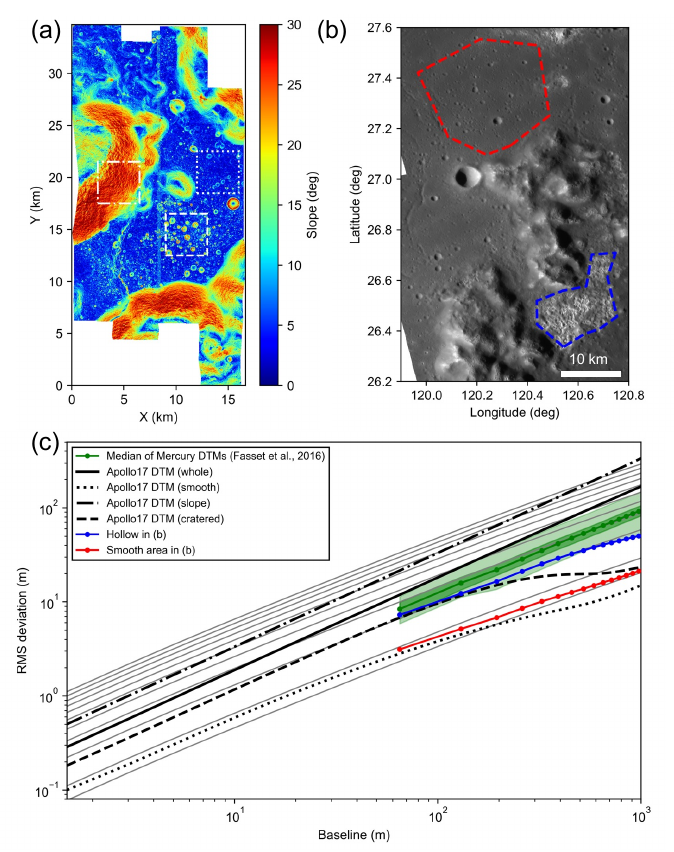}
    \caption{Comparison of topography models. (a) Slope map of the Apollo-17 landing site at a baseline of 1.5 m using a DTM from \citet{haase_coordinates_2019}. The white dotted, dashed, and dash-dotted lines indicate smooth, cratered, and sloped areas used in the roughness analysis in (c), respectively. (b) MDIS image of a hollow area on Mercury. Areas outlined by red and blue dashed lines denote smooth and hollow terrains featured in our roughness analyses in (c). (c) Comparison of RMS deviations among our analog DTMs. The black lines show the roughness of the Apollo-17 DTM, with the solid, dotted, dashed, and dash-dotted lines representing the averages for the entire DTM, smooth, cratered, and sloped areas, respectively. Note that the sloped areas show highest roughness because the RMS deviation is dominated by slope rather than surface curvatures. The green points indicate the median RMS deviations of Mercury DTMs from \citet{fassett_ames_2016}, while thin and thick green shaded regions denote the 1--99\% and 25--75\% quantile ranges, respectively. The blue and red points represent RMS deviations for hollow and smooth terrains on Mercury shown in (b). The gray lines correspond to RMS deviations of the 11 synthetic fractal DTMs used in this study.}
    \label{fig:roughness_comparison}
\end{figure}

Our analysis of RMS deviations of the analog DTMs indicates that the roughness range of these DTMs covers the range present on Mercury's surface. Figure \ref{fig:roughness_comparison}c compares the RMS deviations derived from all analog DTMs employed in this study and the Mercury DTMs from \citet{fassett_ames_2016}. Due to the limited resolution of existing Mercury DTMs, the calculation of Mercury’s roughness is restricted to baselines down to 65 m. At all analyzed baselines, the RMS deviations of Mercury are consistently lower than those of the Apollo-17 DTM when averaged over the entire area. Specifically, 99\% of the RMS deviations derived from Mercury DTMs created by \citet{fassett_ames_2016} fall below the RMS deviation of the whole Apollo-17 DTM, confirming that the Apollo-17 DTM may be an upper bound on the Mercury roughness. Even when spatial variability of roughness is considered, the analog DTMs cover the entire range of Mercury's surface roughness. Our analysis of a DTM containing one of the largest hollows reported by \citet{fassett_ames_2016} (Figure \ref{fig:roughness_comparison}b) shows that hollow interiors are rougher than the adjacent smooth terrain, primarily due to the presence of topographic depressions at baselines of several hundred meters. Although roughness at baselines of several tens of meters remains uncertain for Mercury, the range of roughness variations in the Apollo-17 DTM is likely comparable to those associated with specific geomorphological features on Mercury. Furthermore, the synthesized fractal topographies cover the full range of RMS deviations observed in Mercury DTMs from \citet{fassett_ames_2016}, implying that our analog DTMs may represent Mercury topographic variations.\\

\section{BELA measurement model}
\label{sec:model}
The principle of BELA is to transmit a laser pulse toward the surface of Mercury and to receive the reflected laser pulse, whose travel time indicates a two-way range between MPO and the planetary surface. Prior to the detection of the return pulse, the laser pulse signal undergoes the following four processes that modify its shape from Gaussian assumptions: pulse transmission (section \ref{subsec:emission}), surface reflection and modification by topography within the laser footprint (section \ref{subsec:topography effect}), pulse reception and analog bandpass filtering (section \ref{subsec:filtering}), and noise contamination (section \ref{section:noise}). After accounting for all the four processes, our performance simulation finally conducts the onboard detection algorithm of BELA and generates CW data that represents future observations. The instrument parameters used in this study are summarized in Table \ref{tab:parameters}. All abbreviations are summarized in Section "List of abbreviations".\\

\begin{table}[ht]
\centering
\caption{Instrument parameters of BELA}
\begin{tabular}{|l c c c|}
\hline
Parameter   & Symbol & Value & Unit \\
\hline
Transmitted pulse energy & $E_T$ & 50 & mJ \\
Laser wavelength & $\lambda_T$ & 1064 & nm \\
Telescope radius & $r_R$ & 10 & cm \\
Field of view (half cone) & $\theta_{FOV}$ & 247.5 & $\mu$rad \\
Optical efficiency & $\epsilon_0$ & 0.84 & -- \\
Optical filter efficiency & $\epsilon_t$ & 0.85 & -- \\
Quantum efficiency & $\epsilon_{qe}$ & 0.34 & -- \\
Optical filter width & $\sigma_{rf}$ & 2 & nm \\
APD multiplication gain & $M$ & 48.4 & -- \\
Analog TIA bandwidth & $B_0$ & 20 & MHz \\
AEU gain factor (at gain code 4) & $G_{AEU}$ & 5.19 & -- \\
ADC sampling interval & $\Delta t_{ADC}$ & 12.5 & ns \\
\hline
\end{tabular}
\label{tab:parameters}
\end{table}

\subsection{Pulse transmission}
\label{subsec:emission}

The laser pulse transmitted by BELA exhibits a non-Gaussian profile in both the temporal and spatial domains. The total number of transmitted photons is given by:
\begin{linenomath}
\begin{align}
N_0 = \frac{E_T \lambda_T}{hc},
\label{equation:total_energy}
\end{align}
\end{linenomath}
where $E_T$ is the pulse energy, $\lambda_T$ is the laser wavelength, $h$ is Planck’s constant, and $c$ is the speed of light. The photon emission rate varies temporally, producing a non-Gaussian pulse shape in time as shown in laboratory measurements of the BELA transmitter by \citet{althaus_bela_2019}. In addition, the laser beam diverges spatially as it propagates toward the planetary surface. While previous performance models assumed a two-dimensional Gaussian spatial distribution, laboratory measurements demonstrated that the laboratory-measured beam profile deviates from an ideal Gaussian shape \citep{althaus_bela_2019}. \\

In this study, we incorporate the laboratory-measured temporal pulse shape and spatial beam profile reported by \citet{althaus_bela_2019} to model realistic laser pulse transmitted by BELA as:
\begin{linenomath}
\begin{align}
n_T(t,x,y) = N_0 S_T(t) D_T(x,y),
\label{eq:spatial_distribution}
\end{align}
\end{linenomath}
where $S_T(t)$ is the normalized temporal pulse shape, and $D_T(x,y)$ is the normalized spatial photon distribution (Figure \ref{fig:schematic_method}a). Note that both functions are normalized by the total energy. Although \citet{althaus_bela_2019} showed that the temporal and spatial beam profiles depend on the thermal environment of BELA, we employ the data measured under the "ambient" temperature condition defined in their work. Under other temperature conditions, the return-pulse shape is not expected to differ substantially from the ambient case (see Appendix \ref{sec:pulse_temperature}). \\

\subsection{Modification by topography within footprints}
\label{subsec:topography effect}
The transmitted photons described by Eq. \ref{eq:spatial_distribution} are scattered and reflected by the surface within the laser footprint according to its photometric properties. Following the formulation by \citet{yamada_derivation_2022}, the number of reflected photons per time and solid angle at each surface element, $n_R(t,x,y)$ is:
\begin{linenomath}
\begin{align}
n_R(t,x,y) = n_T(t+\Delta t(x,y),x,y)\,\frac{\alpha_N}{\pi}\,\xi(x,y),
\label{eq:photon_basic}
\end{align}
\end{linenomath}
where $\alpha_N$ is the reflectance at 1064-nm wavelength, $\Delta t(x,y)$ is the two-way travel time of a laser pulse between MPO and the surface element, and $\xi(x,y)$ is the radiation factor describing the surface photometric behavior. In this study, $\xi(x,y)$ is fixed to unity, corresponding to the Lommel-Seeliger reflection law. The Lommel-Seeliger law has been known to approximate the reflectance behavior of airless planetary surfaces and has been widely applied to photometric corrections of reflectance data \citep[e.g.,][]{besse_visible_2013,yamada_derivation_2022}. As proposed by Barron et al. (submitted to this issue), more realistic reflection law may be incorporated in the future.
The travel time $\Delta t(x,y)$ is given by:
\begin{linenomath}
\begin{align}
\Delta t(x,y) = \frac{2(H - z(x,y))}{c},
\label{eq:time_delay}
\end{align}
\end{linenomath}
where $H$ is the MPO altitude above the reference surface of Mercury, and $z(x,y)$ represents the topographic height at the surface element relative to $H$.\\

The number of photons received by BELA is obtained by integrating $n_R(t,x,y)$ over the field of view (FOV) and the telescope aperture. Because the BELA footprint always lies within the FOV \citep{althaus_bela_2019,gouman_measurement_2014}, the received photon flux is proportional to the solid angle of the receiver aperture, $\Omega_R$. Using the telescope radius $r_R$, $\Omega_R$ is given by:
\begin{linenomath}
\begin{align}
\theta_{AP} &= \tan^{-1}{\left(\frac{r_R}{H}\right)},\\
\Omega_R &= 4\pi \sin^2{\left(\frac{\theta_{AP}}{2}\right)},
\label{eq:solid_angle}
\end{align}
\end{linenomath}
where $\theta_{AP}$ is the aperture of the BELA telescope seen from the Mercury surface. The photon rate incident on the avalanche photodiode (APD) of BELA is then:
\begin{linenomath}
\begin{align}
n^{APD}_R(t) = \iint_{\mathrm{FOV}} n_R(t,x,y)\,\Omega_R\,\epsilon_0\,\epsilon_t\,dx\,dy,
\label{eq:received_apd}
\end{align}
\end{linenomath}
where $\epsilon_0$ is the optical efficiency of the telescope, and $\epsilon_t$ is the transmission efficiency of the optical bandpass filter. The corresponding optical power is:
\begin{linenomath}
\begin{align}
P_R(t) = \frac{hc}{\lambda_T}\,n^{APD}_R(t).
\end{align}
\end{linenomath}

As illustrated in Figures \ref{fig:schematic_method}a and b, spatial variations in surface topography within footprints introduce modifications of the temporal shape of return pulses through the travel-time delay. For example, when a boulder is present within a laser footprint, photons reflected from the boulder arrive at BELA earlier than those reflected from the surrounding terrain, producing a multi-peaked return pulse. Additional examples and validation against analytical solutions are provided in Appendix \ref{sec:pulseshape_detail}. \\

\subsection{Pulse reception and analog filtering}
\label{subsec:filtering}

The photon arrival rate is converted into a photocurrent by the APD and subsequently amplified by the Avalanche Photodiode Assembly (APD-A) and the Analogue Electronics Unit (AEU). The photocurrent generated by the APD is:
\begin{linenomath}
\begin{align}
I(t) = q \epsilon_{qe} M n^{APD}_R(t),
\end{align}
\end{linenomath}
where $q$ is the elementary charge, $\epsilon_{qe}$ is the APD quantum efficiency, and $M$ is the APD multiplication gain. $M$ is dependent on the APD bias voltage $V_{bias}$, which accelerates the electrons in the APD. The dependency can be expressed as follows using Miller’s equation \citep{miller_avalanche_1955}:
\begin{linenomath}
\begin{align}
M = \frac{1}{1-\left(\frac{V_{bias}}{V_{br}}\right)^{0.103624}},
\end{align}
\end{linenomath}
where $V_{br}$ is the breakdown voltage of the APD. As shown by \citet{thomas_bepicolombo_2021}, the breakdown voltage depends on the APD temperature. Although the APD bias voltage could be increased to the bias voltage of 370 V \citep{thomas_bepicolombo_2021}, corresponding to $M \approx 113$, the nominal operating plan uses a bias voltage of 320 V ($M = 48.4$) to minimize the risk of APD breakdown. We therefore use this nominal gain value in our simulations to provide conservative performance estimates. \\

The photocurrent is further amplified by the AEU, whose gain is controlled by an integer gain code between 0 and 15. Analysis of dark-noise data obtained during in-cruise checkout and pre-launch testing indicates that the signal-to-noise ratio reaches a minimum at gain code of 4 \citep{stark_-cruise_2024}. Accordingly, we fix the gain code to 4 in all simulations, corresponding to an amplification factor of $G_{AEU}=5.19$ (Appendix \ref{sec:butterworth}). The amplified signal is converted to voltage units as:
\begin{linenomath}
\begin{align}
V(t) = G_{AEU} I(t) R_f ,
\end{align}
\end{linenomath}
where $R_f$ is the resistance of the feedback resistor within the transimpedance amplifier (TIA). \\

Through the amplification process, the limited bandwidth in the receiver electronics introduces additional broadening of the pulse shape. Previous studies approximated this effect using a Gaussian convolution with an idealized digital filter with a bandwidth $B_0$ \citep{steinbrugge_performance_2018,hosseiniarani_comprehensive_2021}. The pulse-broadening effect $\sigma_b$ of the filter is denoted as:
\begin{linenomath}
\begin{align}
\sigma_b = \frac{1}{2\sqrt{2\pi}B_0}\,.
\end{align}
\end{linenomath}
However, the actual frequency response of the BELA receiver is better described by a third-order Butterworth filter \citep{thomas_bepicolombo_2021}. In this work, we model the analog filtering using a Butterworth transfer function whose cut-off frequencies are fitted to laboratory measurements of the BELA Ground Reference Model (GRM) receiver response (see Appendix \ref{sec:butterworth}). \\

The filtered analog signal is sampled and digitized using two phase-shifted analog-to-digital converters (ADCs), yielding an effective sampling rate of 80 MHz (points in Figure \ref{fig:schematic_method}c). The digitized signal is also contaminated by noise (section \ref{section:noise}) and inserted into a range window consisting of 16,064 samples to simulate onboard detection by the Range Finder Module (RFM). Through the detection with the digital matching, the pulse arrival times, widths, amplitudes are computed for four candidates. After computing the travel times of the four candidate pulses identified within the range window, the analysis uses a candidate for which the difference from the expected travel time is minimal. Since this study focuses on the range error introduced by non-Gaussian pulse shapes, the false detections are eliminated from the datasets by comparing the range measurements of the four candidates to the true range. Repeating this process with various surface reflectance values, noise levels, and MPO altitudes, the return pulse signal and onboard-fitting parameters are analyzed to evaluate the BELA performance.

\subsection{Noise model}
\label{section:noise}
\begin{figure}[ht]
\centering
    \includegraphics[width=1\linewidth]{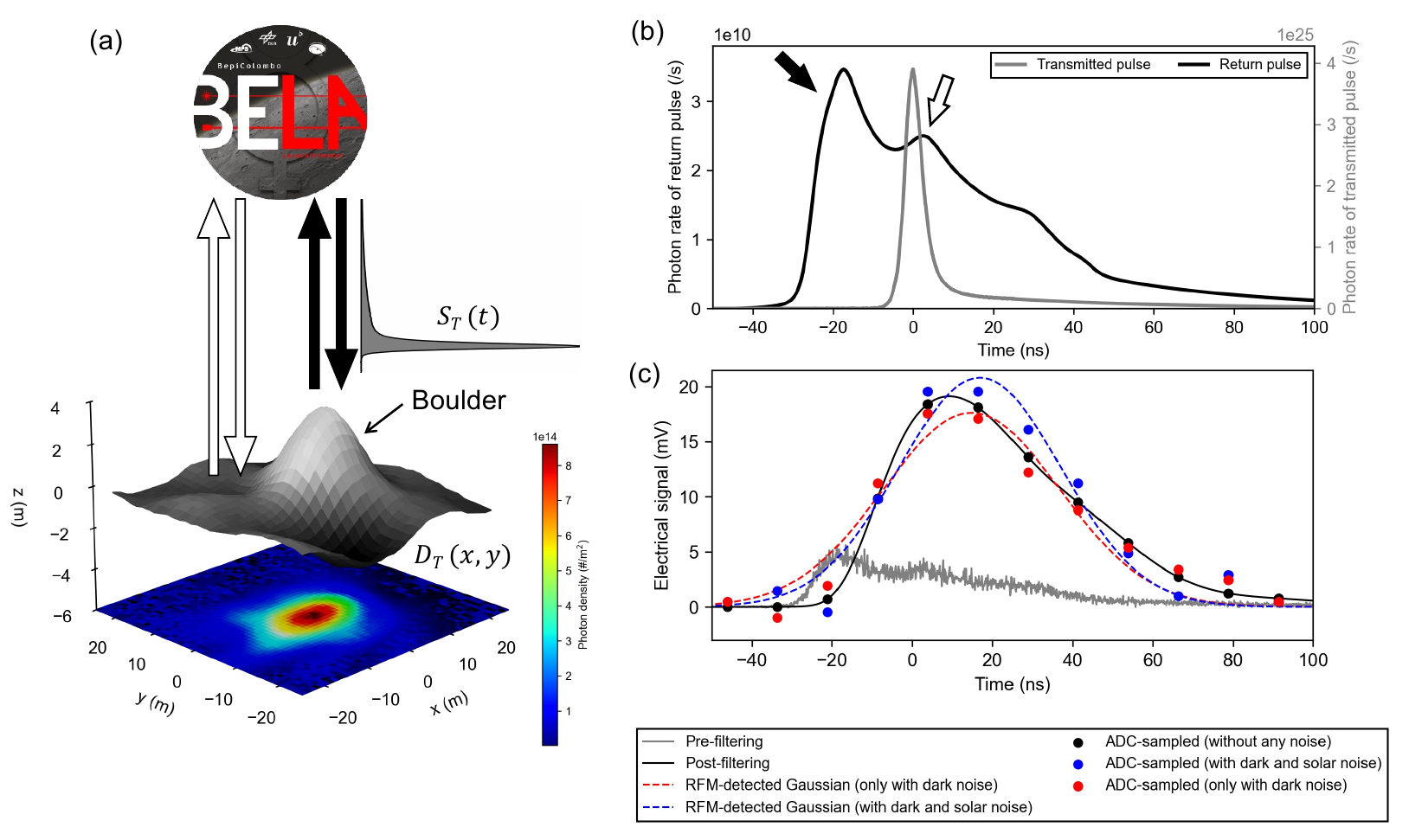}
    \caption{Example of pulse-shape simulation. (a) Schematic illustration of travel time of a laser pulse between MPO and a typical lunar boulder included in the Apollo-17 DTM. (b) Photon rates transmitted by BELA and received by the APD. Return pulse peaks featured by black and white arrows correspond to reflections from the boulder and surrounding terrain shown in (a). (c) Electrical signal processing and pulse detection in the RFM. The gray curve represents the analog voltage signal before the amplification process in the AEU. The black curve shows the filtered and amplified signal. The colored points denote ADC-sampled data with different noise contributions. The colored dashed curves indicate Gaussian fits performed onboard by the RFM.}
\label{fig:schematic_method}
\end{figure}

Noise in BELA measurements arises from three primary sources: shot noise, dark noise, and solar noise. The shot noise originates from the stochastic nature of the multiplication process in the APD and depends on the incident power. The standard deviation of the shot noise in volt is expressed as:
\begin{linenomath}
\begin{align}
\delta V_{\mathrm{shot}}(t) = \left(2q^2\epsilon_{qe}I(t)M^{2+x}B_0R_f^2\right)^{1/2},
\end{align}
\end{linenomath}
where $x$ is the excess noise exponent. To simulate this effect, a random noise array with the above standard deviation is added to $I(t)$ prior to analog filtering (gray curve in Figure \ref{fig:schematic_method}c). \\

The dark noise primarily originates from surface and bulk dark currents in the APD as well as electronic noise in the receiver system. During the cruise checkouts, BELA was unable to transmit laser pulses because the obstruction of its FOV by the Mercury Transfer Module (MTM). Instead, the receiver system was activated to record dark noise signals, yielding range window samples composed entirely of dark noise \citep{stark_-cruise_2024}. However, this dark noise level may be lower than the background noise during laser operation. In the pre-launch electromagnetic compatibility (EMC) test, noise caused by laser transmission increased the noise level from the dark noise. Therefore, to evaluate the BELA performance conservatively, our simulation uses dark noise data including the EMC noise. Selecting one of the dark-noise arrays measured during the pre-launch test, the noise data is added to the simulated electrical signal before performing detection by the RFM (red points in Figure \ref{fig:schematic_method}c). \\

The solar noise is produced by sunlight reflected from Mercury’s surface and collected by the BELA telescope. Although the reflected sunlight itself is nearly constant over a single laser repetition interval and does not influence the AC-coupled detection by the RFM, the APD multiplication process introduces a time-varying noise component detectable in the CW samples due to sunlight. The standard deviation of the solar noise in volt, $\delta V_{solar}$, is:
\begin{linenomath}
\begin{align}
A_{FOV} &= \pi H^2\tan^2(\theta_{FOV}),\\
I_{\mathrm{solar}} &= q\epsilon_0\epsilon_t\epsilon_{qe}\alpha_N\cos (i)\,A_{FOV}\Omega_R\frac{F_0\lambda}{hc}\sigma_{rf},\\
\delta V_{\mathrm{solar}} &= \left[2q^2\epsilon_{qe}I_{\mathrm{solar}}M^{2+x}B_0R_f^2\right]^{1/2},
\end{align}
\end{linenomath}
where $A_{FOV}$ is the footprint area of the FOV on the surface, $\theta_{FOV}$ is the half-cone angle of the FOV, $i$ is the solar incidence angle, $F_0$ is the solar flux at Mercury, and $\sigma_{rf}$ is the optical filter bandwidth. We consider two end-member cases: maximum-solar-noise scenario with $i=0$ (i.e., noon in local time; denoted as dayside hereafter), and no-solar-noise scenario corresponding to nightside observations. In the former case, the noise level increases by approximately 80\% for the average reflectance of Mercury, whereas in the latter case the noise level is equivalent to that of dark noise alone (Figure \ref{fig:schematic_method}c). We also simulate noise-free observation, aiming to derive calibration lines and to estimate influence solely caused by non-Gaussian shape of return pulses.\\

\section{Range measurement performance}
\subsection{Calibration of range measurements}
\label{sec:arrival_time_calibration}
In addition to the range between the Mercury surface and BELA, the time of flight of BELA laser pulse contains detection delays due to electronics, optics, and receiver chain responses. The causes of this delay can be classified into two groups: fixed offsets and variable offsets. This section summarizes all the factors that need to be taken into account when analysing BELA data.

\subsubsection{Fixed offsets}
\hspace{0pt}\newline
The primary cause of fixed offsets is the delay of the transmitted pulse as it travels through an optical fiber between the transmitter and the APD. Once the transmitter emits a laser pulse, a small fraction of the pulse is extracted at a beam splitter and sent to the APD via an optical cable. As the time of flight is obtained by measuring the arrival time of return pulses with respect to that of transmitted pulses at the APD, the light path between the transmitter and the APD needs to be corrected. Considering the refractive index and length of the optical fiber as well as the free path length to the APD, the time of flight between the transmitter and the APD needs to be added to the measurements.\\

The second offset arises from the reflection of the received laser pulse within the BELA telescope. Once a laser pulse is received by the BELA telescope, the pulse is first reflected at a parabolic primary mirror at the base of the telescope, then reflected again at a hyperbolic secondary mirror, and directed toward the APD. The time of flight for the round trip between these two mirrors needs to be subtracted from the measured value.\\

The third fixed offsets is caused by the distance along the line of sight between the APD and the transmitter. A plane that is centered on the beam splitter and is perpendicular to the line of sight will be used as the reference for range measurements. As the APD is positioned on the target side with respect to the reference plane \citep{thomas_bepicolombo_2021}, this position offset between the APD and the transmitter needs to be added to the measured time-of-flight.\\

All the aforementioned offsets are calculated based on the Computer-Aided Design model of BELA. In total, the time of flight measured by BELA contains a systematic reduction of 2.616 ns, corresponding to the range reduction of 0.784 m. The sum of all offsets needs to be added before deriving the one-way range between Mercury and BELA once the measurements start. Table \ref{tab:tof_correction} summarizes the estimation of fixed offsets in the time-of-flight measurements by BELA.\\

\begin{table}[ht]
\centering
\caption{Fixed offsets in time-of-flight measurements by BELA}
\begin{tabular}{l c c}
\hline
Source   & Time-of-flight offset (ns) & Corresponding range (m) \\
\hline
Travel time between transmitter and APD & +3.407 & +1.021 \\
Additional light path within telescope & -0.936 & -0.281 \\
Distance between APD and transmitter & +0.146 & +0.044 \\
\hline
Sum & +2.616 & +0.784 \\
\hline
\end{tabular}
\label{tab:tof_correction}
\end{table}

\subsubsection{Variable offsets}
\label{sec:arrival_time_bias}
\hspace{0pt}\newline
Another bias in the time-of-flight measurements is variable offsets induced due to the modification of return pulse shapes by the analog bandpass filter. In Figure \ref{fig:schematic_method}c, the peak of the pre-filtering pulse is delayed by more than 20 ns after the filtering process. This time shift is caused by the finite impulse response of the frequency-dependent Butterworth filter; therefore, the arrival-time shift depends on the shape of the return pulse. Because both transmitted and received pulses are detected by the RFM, a systematic difference in arrival time between narrow and wide pulses may introduce a second-order bias in range measurements, requiring an additional correction. The delay in the transmitted pulse detection can be assumed to be constant because the temporal shape of the transmitted pulse does not change drastically (see more details in Appendix \ref{sec:tx_delay}). However, the delay in the received pulse should be treated differently as its temporal shape becomes different in every measurement.\\

\begin{figure}[ht]
\centering
    \includegraphics[width=1\linewidth]{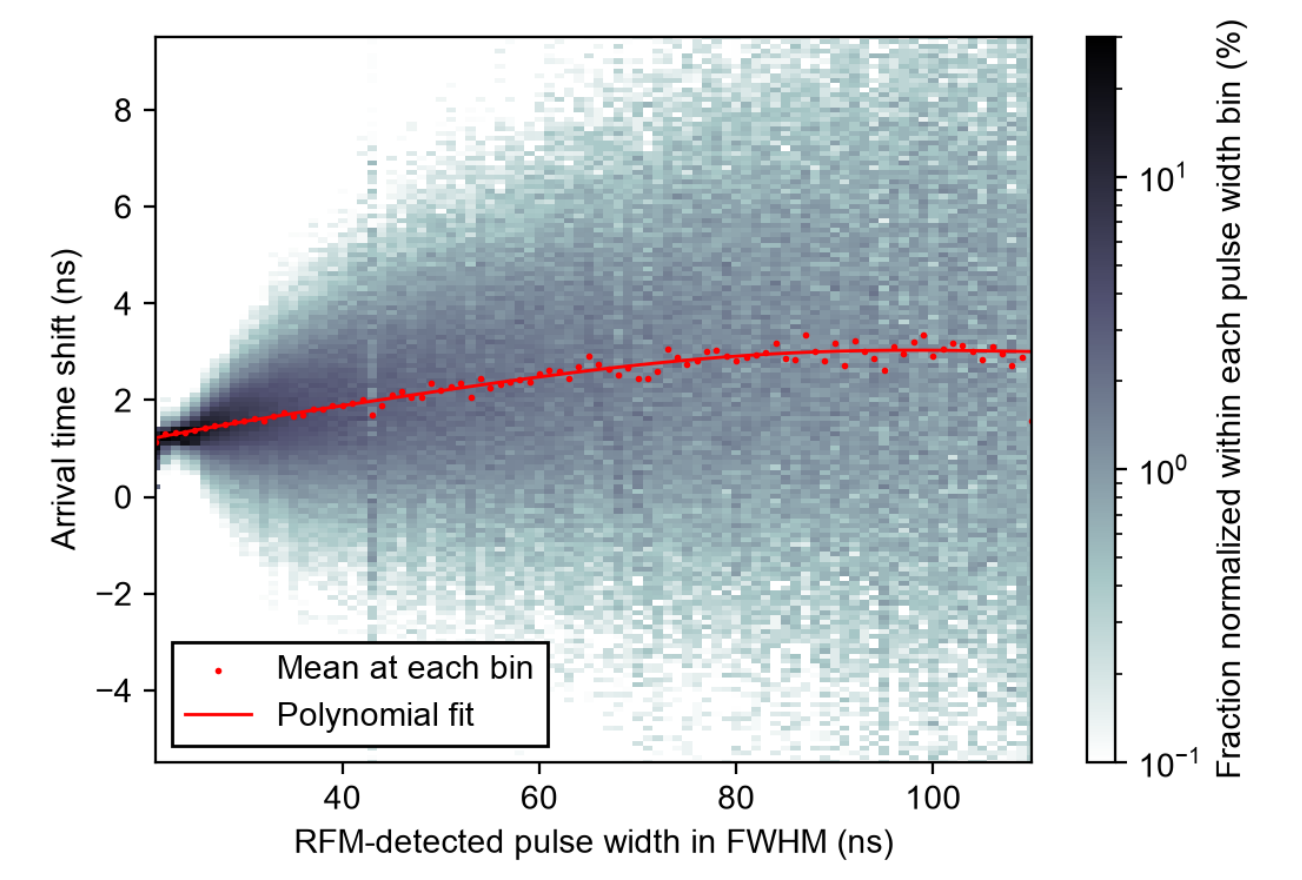}
    \caption{Arrival-time shift due to non-Gaussian return pulses based on our simulations using the Apollo-17 DTM. The background colormap shows the distribution of arrival-time shifts at each pulse width detected by the RFM. Note that the distribution is normalized within each pulse-width bin. The red points indicate the mean arrival-time shift in each bin. The red curve shows a polynomial fit to the mean values.}
\label{fig:time_shift}
\end{figure}

A correction curve for the arrival-time shift can be derived as a function of pulse width, which primarily controls the dominant frequency of the pulse shape. Figure \ref{fig:time_shift} shows the variation in arrival-time shift caused by non-Gaussian return pulse shapes as a function of the RFM-detected pulse width, defined as the full width at half maximum (FWHM), using different altitudes for the Apollo-17 DTM. The arrival-time shift is not uniform at a given pulse width because of variations introduced by non-Gaussian pulse shapes. Note that the influence of the time shift of the transmitted pulse is corrected in Figure \ref{fig:time_shift}. By taking the mean arrival-time shift in each pulse-width bin, a correction curve can be characterized by the following polynomial fit:
\begin{linenomath}
\begin{align}
t_{\mathrm{shift}} = 1.01\times10^{-9} W_G^5
-3.27\times10^{-7} W_G^4
+3.81\times10^{-5} W_G^3-2.18\times10^{-3} W_G^2
+9.37\times10^{-1} W_G
-8.13\quad [\mathrm{ns}],
\end{align}
\end{linenomath}
where $W_G$ is the RFM-detected FWHM in ns. Note that this formula is applicable to pulses with RFM-detected FWHMs of 20--110 ns, corresponding to FWHMs detectable in the nominal RFM configuration.\\

This correction can reduce systematic range errors by decimeter scales. Compared to the FWHM of the transmitted pulse detected by the RFM, the detected arrival time of pulses with widths of approximately 100 ns may have a bias of about 3 ns, corresponding to a one-way range error of 0.5 m. Note that all range values reported later refer to one-way range measurements. Variations from the correction curve remain after applying the offset correction and contribute residual range errors with a standard deviation of up to 1 m.

\subsection{Range errors}
\label{sec:range_results}
After applying the arrival-time offset correction, our numerical simulations provide updated and detailed estimates of BELA range errors compared to previous studies by \citet{steinbrugge_performance_2018} and \citet{hosseiniarani_comprehensive_2021}. Range errors in BELA measurements arise from three primary sources. First, non-Gaussian return pulse shapes introduce errors in determining the Gaussian center during onboard fitting processes. Although systematic biases can be mitigated using pulse-width-based corrections (section \ref{sec:arrival_time_bias}), residual errors remain and contribute to the overall range uncertainty. Second, deviations in the pointing direction introduce additional range errors. Laboratory measurements by \citet{althaus_bela_2019} show that the pointing direction of the transmitted pulse varies from shot to shot, even under the same temperature conditions. When the illuminated surface has a non-zero slope, these pointing variations contribute to variations in ranges between the footprint and spacecraft. Third, electrical noise in the received signal further degrades range measurement accuracy. Because solar background noise alters the total noise level, range measurement performance depends on the local time of the footprint. In this study, we treat models with and without solar noise as representative of dayside and nightside measurements on Mercury, respectively. \\

\begin{figure}[ht]
\centering
    \includegraphics[width=1\linewidth]{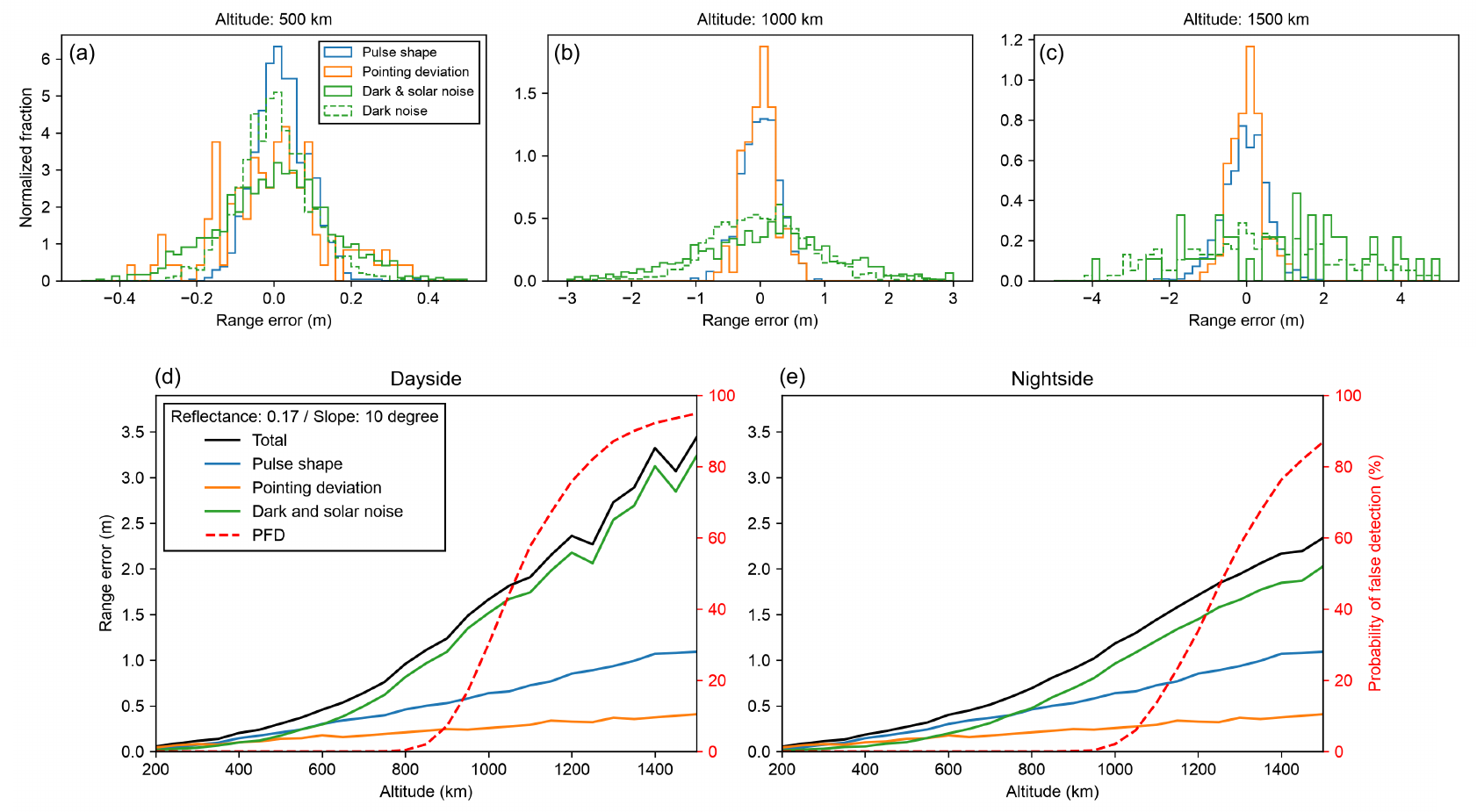}
\caption{Range errors from various sources. (a–c) Histograms of range errors from individual error sources at altitudes of 500, 1000, and 1500 km, respectively. (d) Range errors in dayside as a function of MPO altitudes. The red dashed line shows the probability of false detections (PFD). The black lines denote the total range error from all the noise sources. (e) Same as (d) but in nightside.}
\label{fig:range_errors}
\end{figure}

Figures \ref{fig:range_errors}a–c show distributions of range errors from each error source at altitudes of 500, 1000, and 1500 km. In these Figures, the 1064-nm surface reflectance is assumed to be 0.17, corresponding to the mean value derived from MLA measurements \citep{neumann_bright_2013}. The results are shown for a DTM slope of 10 degrees, which represents the expected average slope at the BELA footprint scale, based on extrapolation of average slope from kilometric to tens-of-meter baselines by \citet{hosseiniarani_comprehensive_2021}. The DTM slope is estimated by fitting a plane to elevations within each footprint. At an altitude of 500 km, all three error sources contribute comparably to the total range error, highlighting the importance of accounting for each error component. At altitudes of 1000 and 1500 km, range errors are dominated by dark and solar noise. The resulting range errors show approximately normal distributions; therefore, the standard deviation is used as a metric of range errors in the following analyses.\\

Figures \ref{fig:range_errors}d and e show the standard deviation of range errors from each source as a function of MPO altitude, assuming a slope of 10 degrees and an reflectance of 0.17. All error components increase with altitude. In particular, noise-related errors dominate above altitudes of approximately 800 km, because noise effects depend on the signal-to-noise ratio, which decreases with increasing the square of distance (Eqs. \ref{eq:solid_angle}-\ref{eq:received_apd}). At 1400 km, corresponding to the maximum altitude for BELA operation, the total standard deviation of the range error reaches approximately 3 m on dayside and 2 m on nightside. In this study, detected return pulses with range errors exceeding 10 m are classified as false detections and excluded from the range-error statistics. This exclusion also allows estimation of the probability of false detection (PFD) as a function of altitude. For nominal Mercury conditions (slope of 10 degrees and reflectance of 0.17), the PFD at 1400-km altitude is estimated to be 92\% on the dayside and 76\% on the nightside, indicating that pulse detection remains possible with probability of 8–24 \% throughout the entire orbit.\\

\begin{figure}[ht]
  \centering
      \includegraphics[width=1\linewidth]{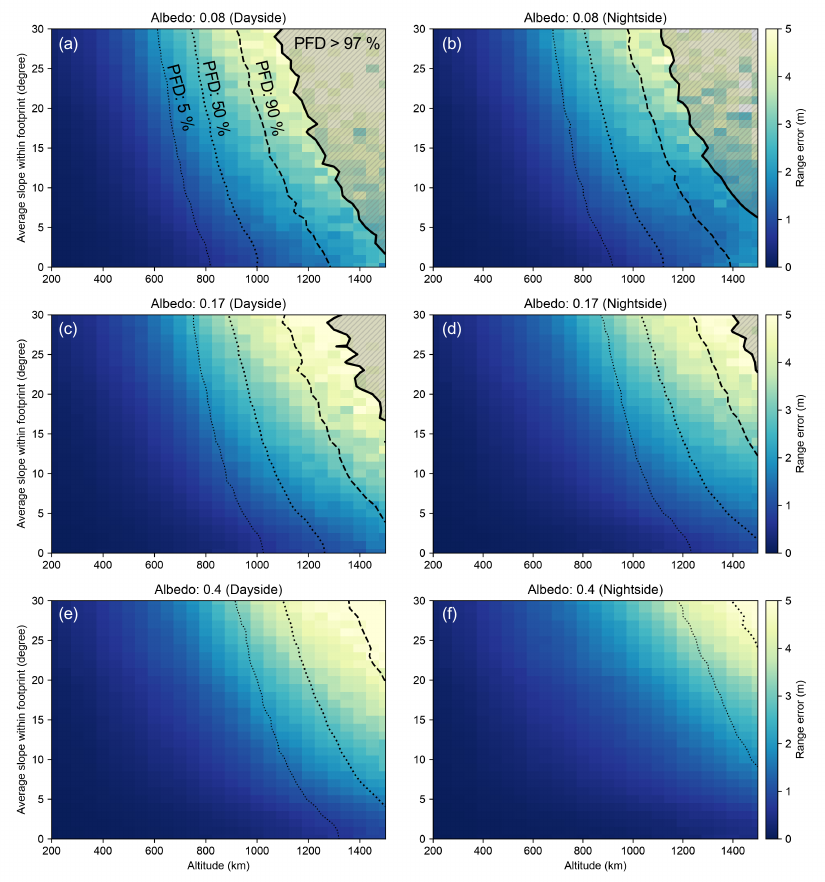}
  \caption{Total range errors under dayside and nightside conditions for different combinations of surface slope, MPO altitudes, and reflectance. The upper, middle, and bottom panels show results with typical reflectance of dark deposits, ice-free regolith, and ice exposures in PSRs according to \citet{barker_new_2022}. In all panels, the gray shade shows parameter sets with PFD exceeding 97 \%. The thin dotted, thick dotted, dashed, and solid black lines show contours at PFDs of 5, 50, 90, and 97 \%, respectively.}
  \label{fig:range_error_map}
  \end{figure}

The range error and PFD are strongly dependent on surface reflectance and slope within the footprint, as both parameters affect the amplitude of the return pulse. Figure \ref{fig:range_error_map} shows range errors for various slopes and altitudes, assuming typical reflectance for dark deposits, ice-free regolith, and icy permanently shadowed regions (PSRs) as reported by \citet{barker_new_2022}. The simulations indicate that, when the normal regolith surface is observed by BELA, even steep slopes of 30 degrees can be detected at altitudes below 1100 km without the PFD exceeding 90\% (Figures \ref{fig:range_error_map}c and d). The PFD remains below 5\% up to altitudes of 800 km. Given that lunar slopes at a 25-m baseline are always less than 30 degrees \citep{kreslavsky_steepest_2016}, and that Mercury is expected to exhibit gentler slopes, BELA should be capable of adequately capturing Mercury’s topographic profiles on average. Even in PSRs, where dark deposits and/or ice materials may be present \citep[e.g.,][]{chabot_images_2014}, reliable range measurements are achievable below altitudes of 800 km. For dark deposits (reflectance of $\sim$0.08), range measurements may remain feasible up to altitudes of approximately 1000 km if the surface of such deposits is not inclined. Table \ref{tab:typical_range_errors} summarizes typical values of the estimated range errors.

\begin{table}[ht]
\centering
\caption{Estimated range errors and PFDs in BELA measurements at a slope of 10 degrees.}
\begin{tabular}{c c c c c}
\hline
\multirow{2}{*}{MPO altitude} & \multirow{2}{*}{Condition} & \multicolumn{3}{c}{Range error (PFD)} \\
\cline{3-5}
& & Dark deposit & Ice-free regolith & Ice exposure \\
\hline
\multirow{2}{*}{500 km} & Dayside & 0.25 m (0.0\%) & 0.20 m (0.0\%) & 0.17 m (0.0\%)\\
& Nightside & 0.22 m (0.0\%) & 0.17 m (0.0\%) & 0.15 m (0.0\%)\\
\hline
\multirow{2}{*}{800 km} & Dayside & 1.52 m (30.8\%) & 1.11 m (0.6\%) & 0.78 m (0.0\%)\\
& Nightside & 1.20 m (8.7\%) & 0.81 m (0.0\%) & 0.65 m (0.0\%)\\
\hline
\multirow{2}{*}{1100 km} & Dayside & 3.68 m (93.8\%) & 3.14 m (76.4\%) & 2.49 m (18.5\%)\\
& Nightside & 2.60 m (91.4\%) & 2.40 m (34.8\%) & 1.65 m (0.0\%)\\
\hline
\multirow{2}{*}{1400 km} & Dayside & 5.61 m (99.0\%) &5.47 m (97.4\%) & 4.86 m (87.9\%)\\
& Nightside & 4.66 m (99.6\%) & 4.83 m (94.9\%) & 4.09 m (25.2\%)\\
\hline
\end{tabular}
\label{tab:typical_range_errors}
\end{table}

\section{Accuracy of surface reflectance measurements}
\subsection{Conversion from electrical signals to return pulse energy}
\label{sec:energy_calibration}
To assess the surface reflectance of Mercury at 1064 nm, the return pulse energy needs to be estimated from the integration of electrical signals of return pulses detected by the RFM using two different approaches. The first approach is to sum CW samples with a trapezoidal integration:
\begin{linenomath}
\begin{align}
E_{CW} = \sum_{k=1}^{41} \frac{V_{CW}[k] + V_{CW}[k+1]}{2}  \Delta t_{ADC} \quad [\text{ns}\cdot\text{mV}],
\end{align}
\end{linenomath}
where $V_{CW}[k]$ denotes the CW samples in units of mV, and $\Delta t_{ADC}$ is the ADC sampling interval. The second approach is to integrate the fitted Gaussian pulse:
\begin{linenomath}
\begin{align}
E_{G} = \frac{\sqrt{\pi}}{2\sqrt{\ln{2}}} W_G A_{G} \quad [\text{ns}\cdot\text{mV}],
\end{align}
\end{linenomath}
where $A_G$ is the amplitude of the Gaussian detected by the RFM in mV.\\

\begin{figure}[ht]
\centering
    \includegraphics[width=0.8\linewidth]{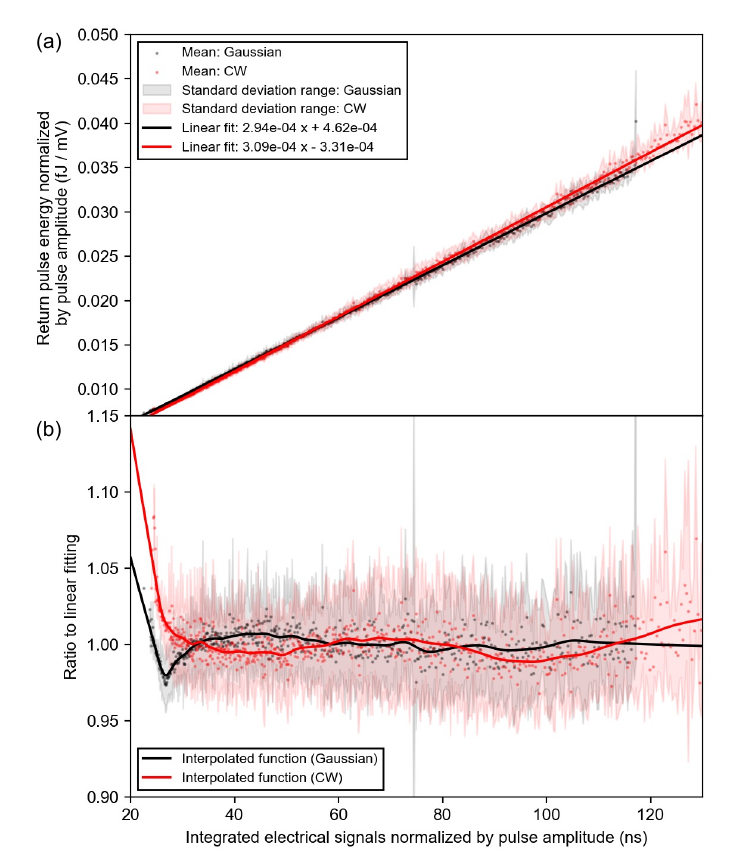}
\caption{Calibration of return pulse energy. (a) Relationship between optical return pulse energy and integrated electrical signals. Both axes are normalized by the detected pulse amplitude. (b) Return pulse energy normalized by the linear fitting results. The points and shaded regions correspond to the same data as in panel (a) after normalization by the linear fits.}
\label{fig:energy_calibration}
\end{figure}

Conversion from the integral of the electrical signal to the actual return pulse energy must account for biases caused by non-Gaussian pulse shapes. Figure \ref{fig:energy_calibration}a shows the relationship between the simulated optical return pulse energy and $E_{CW}$ and $E_G$, after normalization by the detected amplitude, $A_G$, based on noise-free simulations. As a first-order approximation, the comparison reveals a linear relationship between the optical return pulse energy and the integrals of the electrical signals. By fitting linear functions, this relationship can be expressed as:
\begin{linenomath}
\begin{align}
E_{optical}^{G} &=
A_{G}
\left(\frac{48.4}{M}\right)
\left(\frac{5.19}{G_{\mathrm{AEU}}}\right)
\left(2.94\times10^{-4}\frac{E_{G}}{A_{G}} + 4.62\times10^{-4}\right)
\quad [\text{fJ}], \\
E_{optical}^{CW} &=
A_{G}
\left(\frac{48.4}{M}\right)
\left(\frac{5.19}{G_{\mathrm{AEU}}}\right)
\left(3.09\times10^{-4}\frac{E_{CW}}{A_{G}} - 3.31\times10^{-4}\right)
\quad [\text{fJ}],
\end{align}
\end{linenomath}
where $E_{optical}^{G}$ and $E_{optical}^{CW}$ are optical return pulse energies estimated from the Gaussian fitting and CW-sample integrations, respectively. Note that the multiplication gain $M$ and the AEU gain $G_{AEU}$ must be considered according to the observation settings.\\

In addition to this first-order linear relationship, more complex behavior must be considered in the conversion process. Figure \ref{fig:energy_calibration}b shows the return pulse energy normalized by $A_G$ after division by the linear fitting results. The mean ratio relative to the linear fit exhibits a systematic deviation of up to 14 \%, depending on the normalized integral of the electrical signal. This systematic bias arises from the characteristics of the third-order Butterworth filter in the AEU receiver chain. For example, when the return pulse is as narrow as the transmitted pulse, the high-order Butterworth filter produces negative tails and oscillations in electrical signals following the main peak. This systematic bias can be corrected using interpolation functions of the integrated electrical signal, as illustrated in Figure \ref{fig:energy_calibration}b. This procedure reduces the estimation error of the return pulse energy, although complex return pulse shapes caused by within-footprint topography still introduce residual uncertainties at 2.5- and 3.1-\% levels for $E_{optical}^{G}$ and $E_{optical}^{CW}$, respectively.

\subsection{Measurement error of return pulse energy}
Including the noise effect and the calibration described in section \ref{sec:energy_calibration}, the accuracy of return pulse energy measurements is evaluated by comparing the simulated return pulse energy reaching the APD with $E_{optical}^{G}$ and $E_{optical}^{CW}$. Energy estimation is only meaningful for correctly detected pulses; therefore, false detections are excluded using the same criterion applied in the range error analysis.\\

\begin{figure}[ht]
  \centering
    \includegraphics[width=0.7\linewidth]{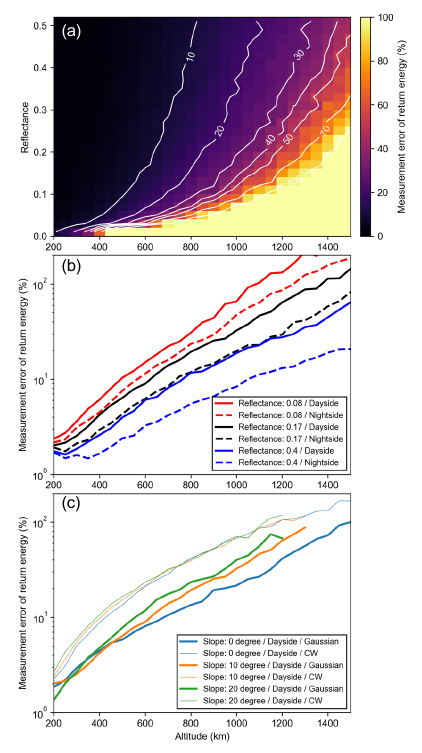}
  \caption{Measurement errors of return pulse energies. (a) Measurement errors at a slope of 10 degrees as a function of altitude and reflectance, using $E_G$ under the dayside noise condition. (b) Measurement errors at a slope of 10 degrees as a function of altitude for different reflectance and noise conditions, using $E_G$. (c) Measurement errors at an reflectance of 0.17 as a function of altitude for different slopes, measurement methods, and noise conditions. Thin and thick lines correspond to estimates based on $E_{CW}$ and $E_G$, respectively. The lines are shown until the PFD reaches 90\%.}
  \label{fig:energy_estimation}
\end{figure}

Figure \ref{fig:energy_estimation}a shows the distribution of estimation errors of the return pulse energy when $E_G$ is used under the dayside noise condition. In this plot, footprints with slopes of 10 degrees are featured. The estimation error is primarily controlled by altitude and surface reflectance as the signal-to-noise ratio is reduced at higher altitudes and lower reflectances. For example, when the surface reflectance is assumed to be the global mean value of Mercury (0.17; \citet{neumann_bright_2013}), the measurement error remains below 30\% at altitudes below 900 km. This accuracy is sufficient to characterize reflectance variations associated with space weathering processes \citep{deutsch_temperature-related_2024} and confirms that reflectance-based geologic characterization in polar regions may be successfully conducted by BELA. Figure \ref{fig:energy_estimation}b provides an overview of energy measurement errors for various combinations of reflectance and noise conditions. Because dark deposits and icy materials are primarily located within PSRs, where solar noise can be neglected, nightside noise conditions are applicable to polar science cases. At altitudes below 900 km, the energy measurement error remains below 30\% for all reflectance scenarios considered. As this error is relative to the reflectance, the corresponding absolute uncertainty is sufficiently small to distinguish dark deposits and icy materials from the average reflectance of ice-free regolith.\\

Our analysis also indicates that the influence of local surface slope on reflectance measurements is limited. Figure \ref{fig:energy_estimation}c shows variations in the energy measurement error for different slopes. Pulse broadening due to increased slope reduces the signal-to-noise ratio and increases uncertainties in pulse amplitude and width estimates, leading to larger energy estimation errors at higher slopes. Nevertheless, results obtained for a slope of 10 degrees represent well the cases with steeper slopes as the estimation error does not change drastically. Furthermore, our simulations reveal that $E_G$ provides more accurate energy estimates than $E_{CW}$, because $E_{CW}$ is more strongly affected by noise during the integration of the CW samples. Therefore, reflectance measurement of BELA needs to be based on fitted Gaussian parameters, rather than use of the CW samples.

\section{Roughness characterization with laser pulse shape}
\subsection{Difficulty of pulse-width usage as a roughness indicator}
\label{subsec:misinterpretation}
Our performance simulations based on the Apollo-17 and fractal DTMs quantitatively demonstrate that conventional roughness estimates derived from return pulse width are misleading when the underlying assumption of random topography is violated. The observed pulse width measured by a laser altimeter has traditionally been formulated as a Gaussian convolution \citep{gardner_ranging_1992}:
\begin{linenomath}
\begin{align}
\label{eq:roughness_ideal}
\sigma_{obs} = \sqrt{\sigma_{trans}^2 + \sigma_{beam-div}^2 + \sigma_{slope}^2 + \sigma_{rough-obs}^2},
\end{align}
\end{linenomath}
where $\sigma_{trans}$ represents the standard deviation of the temporal photon distribution, assuming that the transmitted pulse shape $S_T(t)$ in Eq. \ref{eq:spatial_distribution} is Gaussian. Similarly, $\sigma_{beam-div}$ is defined under the assumption that the spatial beam profile $D_T(x,y)$ follows a two-dimensional Gaussian distribution. $\sigma_{slope}$ is pulse broadening due to within-footprint slopes (see \citet{gardner_ranging_1992} and Appendix \ref{sec:pulseshape_detail}). By subtracting $\sigma_{trans}^2 + \sigma_{beam-div}^2 + \sigma_{slope}^2$ from $\sigma_{obs}^2$, the residual term, $\sigma_{rough-obs}^2$, is interpreted as the contribution from surface roughness. If surface roughness within a footprint can be described as a white-noise-like elevation distribution, the pulse broadening due to roughness is analytically given as:
\begin{linenomath}
\begin{align}
\label{eq:roughness_analytical}
\sigma_{rough-DTM} = \frac{2}{c} \mathrm{Std}(\Delta z(x,y)),
\end{align}
\end{linenomath}
where $\Delta z(x,y)$ denotes the elevation relative to the best-fit plane within the footprint, and $\mathrm{Std}$ stands for standard deviation that defines the footprint-scale roughness as $\mathrm{Std}(\Delta z(x,y))$. Under this theoretical framework, $\sigma_{rough-obs}$ should equal $\sigma_{rough-DTM}$. This assumption has been widely used in previous roughness estimations for the Moon and Mars by analyzing return pulse widths \citep[e.g.,][]{neumann_mars_2003,smith_initial_2010, poole_calibrating_2014}.\\

To examine whether this assumption holds for realistic planetary surfaces in future BELA observations, we compare $\sigma_{rough-obs}$ and $\sigma_{rough-DTM}$ using our numerical simulations. First, we derived the relationship between footprint slope and the RFM-detected pulse width by simulating return pulses from flat surfaces with varying slopes. Applying this relationship to the average slope within each footprint, we then calculate the expected pulse-width contribution from the slope itself at all footprints simulated with the analog DTMs. Next, we subtract the pulse width expected at the same slope angle from the RFM-detected pulse width at each footprint, $\sigma_{obs}$ (defined as $W_G / 2\sqrt{2\ln2}$), to isolate $\sigma_{rough-obs}$. Because $\sigma_{rough-DTM}$ can be directly computed from the DTM elevations within each footprint, we finally compare $\sigma_{rough-obs}$ and $\sigma_{rough-DTM}$.\\

\begin{figure}[ht]
\centering
    \includegraphics[width=0.8\linewidth]{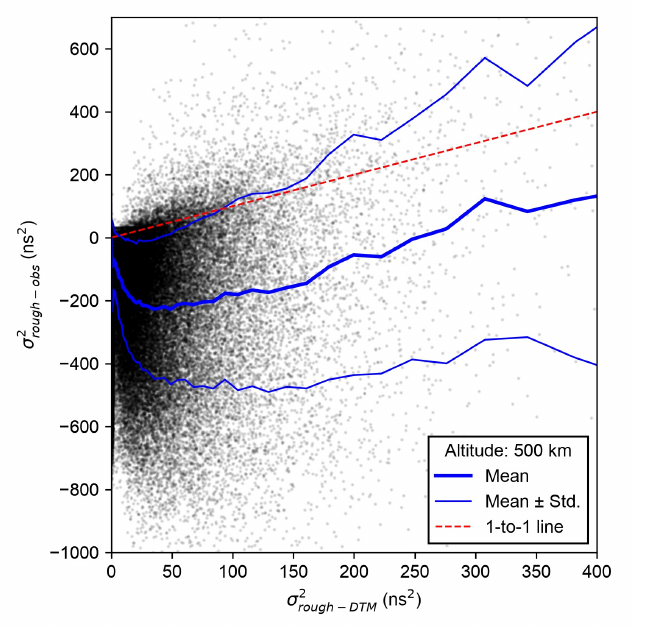}
  \caption{Comparison of $\sigma_{rough-DTM}^2$ and $\sigma_{rough-obs}^2$. Black points represent results from all footprints. The red line indicates the one-to-one relationship, where $\sigma_{rough-DTM}^2$ equals $\sigma_{rough-obs}^2$. The thick and thin blue curves show the mean value and the range of one standard deviation, respectively.}
  \label{fig:pulse_width}
\end{figure}

The comparison reveals that the observed pulse width does not provide a reliable estimate of roughness as formulated in Eq. \ref{eq:roughness_analytical}. Figure \ref{fig:pulse_width} shows the distribution of $\sigma_{rough-DTM}^2$ and $\sigma_{rough-obs}^2$ at an altitude of 500 km using all 11 fractal DTMs. The values of $\sigma_{rough-obs}^2$ exhibit a wide spread even at the same $\sigma_{rough-DTM}^2$ and frequently become negative. Since $\sigma_{rough-DTM}^2$ needs to be positive, these negative values show that a rough surface often generates return pulse narrower than flat surfaces with the same slope angles, indicating a breakdown of the white-noise assumption. Furthermore, averaging $\sigma_{rough-obs}^2$ within roughness bins does not converge to $\sigma_{rough-DTM}^2$, although a weak increasing trend is present.\\

\begin{figure}[ht]
\centering
    \includegraphics[width=0.8\linewidth]{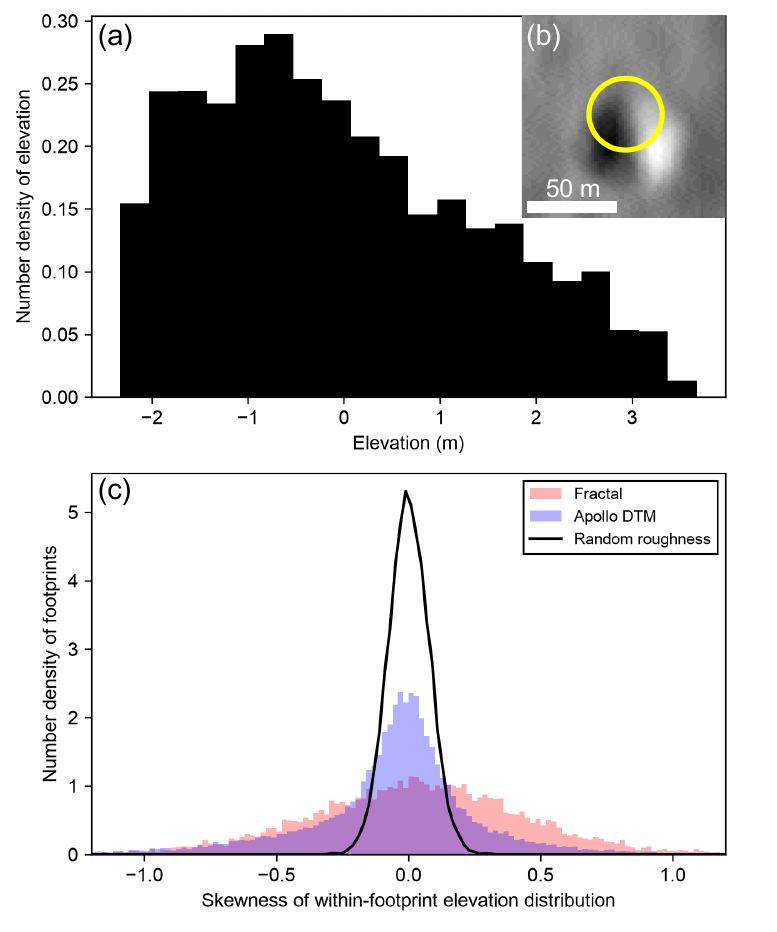}
  \caption{Statistics of topography distributions within footprints. (a) Elevation distribution within a BELA footprint located near a crater with a diamter of 50 m. The altitude is set 500 km. (b) Hillshade map of the area surrounding the footprint analyzed in (a). The footprint is indicated by the yellow circle. (c) Statistics of skewness of elevation distributions for all footprints. The red and blue histograms correspond to results from the Apollo-17 and fractal DTMs, respectively. The black line shows the skewness distribution on an ideally random DTM whose elevation follows a normal distribution like white noise. The Apollo-17 and fractal DTMs have a wider variety of skewness than white-noise-like DTMs.}
  \label{fig:topography_skewness}
\end{figure}

This discrepancy arises from the scale dependence of planetary topography. Eq. \ref{eq:roughness_analytical} implicitly assumes that $\Delta z(x,y)$ is independent of spatial scales. In contrast, real planetary surfaces exhibit scale-dependent roughness, as illustrated in Figure \ref{fig:roughness_comparison}c. As a result, elevation distributions within a footprint are generally non-Gaussian and often skewed. Figures \ref{fig:topography_skewness}a and b show an example of elevation histograms within a footprint located near a crater, where the footprint scale is comparable to the crater diameter. In such cases, the elevation distribution within the footprint is strongly skewed as shown in Figure \ref{fig:topography_skewness}a. We examine skewness of elevation distributions within all 23,292 and 9,801 footprints tested on the Apollo-17 and fractal DTMs, respectively. Statistical analysis of all footprints confirms that the skewness of elevation distributions deviates significantly from that expected for white-noise-like topography (Figure \ref{fig:topography_skewness}c). Compared to a Monte Carlo simulation of idealized random topography (black curve in Figure \ref{fig:topography_skewness}c), real elevation distributions exhibit larger skewness. Therefore, roughness superposed on slopes tends to skew the return pulse shape rather than broaden it symmetrically, introducing biases in Gaussian pulse-width estimates.\\

\subsection{Alternative roughness constraints using pulse shape data}
As demonstrated above, conventional roughness estimates based on $\sigma_{obs}$ are unreliable because of the scale dependency of surface topography. Instead, this study proposes an alternative approach that characterizes the detailed shape of return pulses. A key advantage of BELA is the availability of CW samples, which enables a direct analysis of return pulse shapes. At footprints with significant roughness and non-zero skewness, return pulse shapes become distinctly asymmetric and can be distinguished from pulses reflected from flat but sloped surfaces. Thus, comparison between observed pulse shapes and those expected from ideally flat surfaces may provide information on footprint-scale roughness.\\

To demonstrate this approach, we developed a roughness estimation method using simulated return pulses. First, we generate a library of return pulse shapes reflected from flat slopes. Because the non-axisymmetric spatial beam profile $D_T(x,y)$ makes pulse shapes sensitive to slope azimuth, the pulse library is constructed by varying slope angle and azimuth at 1 degree and 30 degrees intervals, respectively. Next, each simulated return pulse from realistic topography is compared with this pulse library to identify the most correlated flat-slope pulse. After scaling the amplitude of the most correlated pulse, the residual difference between the simulated pulse and the best-fit pulse serves as a proxy for within-footprint roughness.\\

\begin{figure}[ht]
\centering
    \includegraphics[width=1\linewidth]{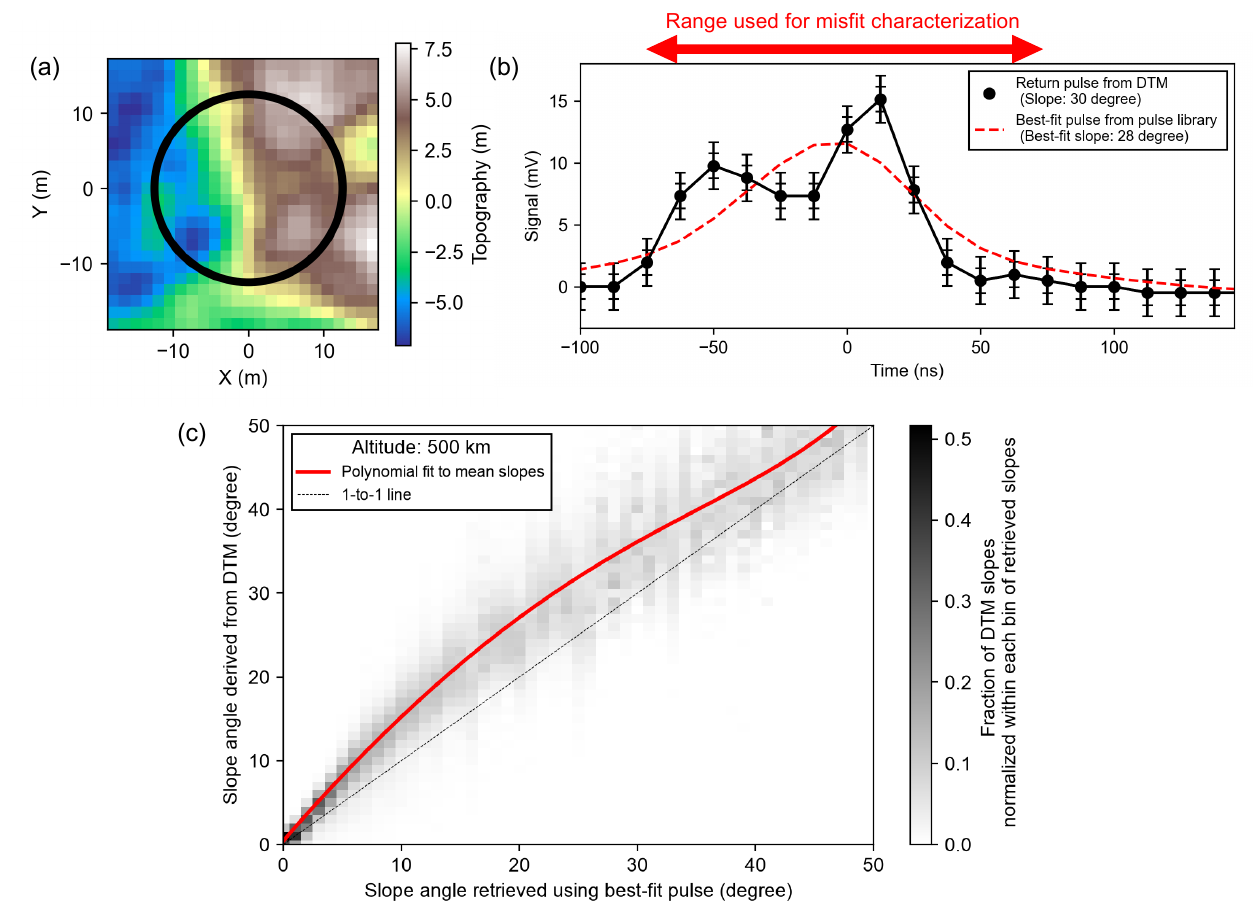}
  \caption{Example of return pulse compared with the pulse library. (a) Topography around a footprint on a highly rough location. The black circle shows the BELA footprint. (b) CW samples at (a) and the best-fit shape found in the flat-slope return pulse library. The return pulse simulated with the DTM and the best-fit pulse selected from the pulse library are shown in black and red, respectively. The large and small black error bars indicate the ranges of noise standard deviations under dayside and nightside conditions, respectively. (c) Comparison of flat slopes derived from the DTM elevations and those retrieved using the best-fit pulse in the pulse library. The background color map represents the normalized fraction of DTM slopes within each best-fit slope bin. The red line shows a polynomial fit to the mean DTM slopes at each bin.}
  \label{fig:pulse_shape_roughness}
\end{figure}

Figure \ref{fig:pulse_shape_roughness} illustrates a footprint example for a highly rough place on the fractal DTM. Due to a cliff-like topography within the footprint, reflections from one side arrive earlier, producing a double-peaked pulse shape (Figure \ref{fig:pulse_shape_roughness}b). The difference between the simulated return pulse and the best-fit flat-slope pulse exceeds the noise level around the pulse peak, with three and seven samples exceeding twice the noise standard deviation under dayside and nightside conditions, respectively. As a secondary outcome, the pulse-library search also provides an estimate of the local slope. However, the slope inferred from the best-fit pulse is systematically lower than that derived directly from DTM elevations (Figure \ref{fig:pulse_shape_roughness}c). In the same way as the bias found in pulse width comparison (section \ref{subsec:misinterpretation}), this systematic bias is caused by the fact that skewed pulses tend to be matched by narrower flat-slope pulses (see examples shown in Appendix \ref{sec:pulseshape_detail}).\\

The residual between the return pulse and the best-fit pulse provides a lower bound on surface roughness. The residual depends on two statistics of the elevation distribution within each footprint: standard deviation (i.e., roughness) and absolute skewness. For example, when both roughness and absolute skewness are high within a footprint, such as at a large boulder, the return pulse deviates significantly from any flat-slope pulse, producing detectable residuals. In contrast, at footprints where absolute skewness is high but roughness is low, such as at very shallow craters, the return pulse shape remains indistinguishable from that from a flat slope. Thus, when using the residual from the best-fit pulse, roughness is detectable only when it exceeds a certain roughness value. Therefore, this method indicates the lower bounds on footprint-scale roughness. In addition, as the absolute skewness can also be as small as zero (Figure \ref{fig:topography_skewness}c), footprints with high roughness can also produce non-skewed return pulses and, hence, are not necessarily detected as anomalies from the best-fit pulse. Therefore, without a priori knowledge of the average slope within the footprint, within-footprint roughness needs to be estimated only in a statistical manner.\\

\begin{figure}[ht]
\centering
    \includegraphics[width=0.8\linewidth]{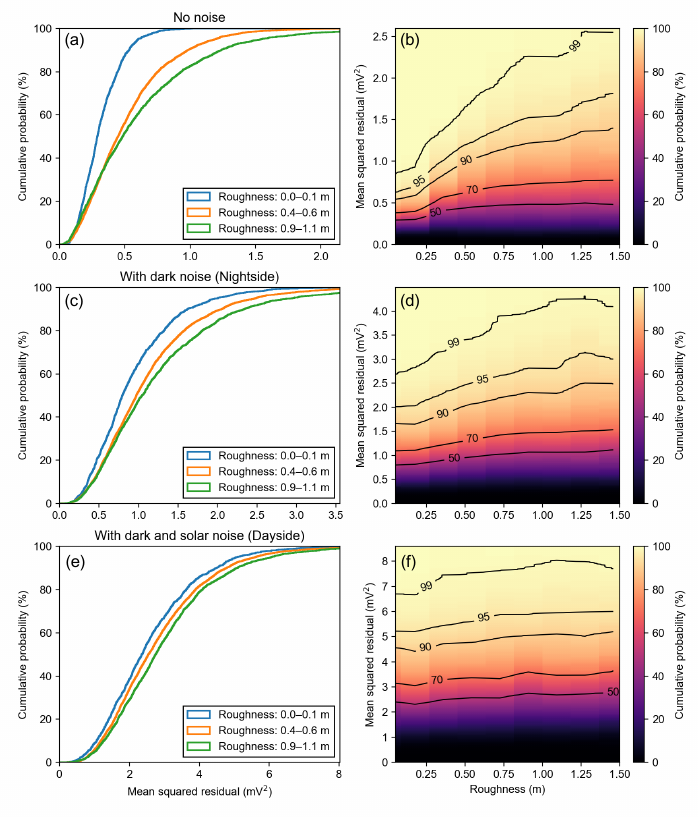}
  \caption{Mean squared residuals (MSRs) of CW samples after subtracting the best-fit pulses found in the pulse shape library. (a, c, e) Cumulative probability of MSRs for three roughness bins. (b, d, f) Cumulative probability of MSRs as a function of surface roughness. The density is normalized within each roughness bin. The black solid lines indicate contours at cumulative probabilities of 50, 70, 90, 95, and 99 \%. (a, b) Results without any noise. (c, d) Results with nightside noise. (e, f) Results with dayside noise. In all panels, the MPO altitude is set 500 km. Only cases with best-fit slope angles of 10 degrees and surface reflectance of 0.17 are included.}
  \label{fig:roughness_constraints}
\end{figure}

Using the mean squared residual (MSR) between the simulated pulse and the best-fit flat-slope pulse, we demonstrate possible constraints on roughness. Subtracting the best-fit pulse retrieved from the pulse library (the red dashed line in Figure \ref{fig:pulse_shape_roughness}b) from the return pulse signals (the black circles in Figure \ref{fig:pulse_shape_roughness}b), the MSR is calculated for all footprints and classified by best-fit slopes and DTM-derived roughness. Figure \ref{fig:roughness_constraints} shows the normalized cumulative probability of the MSR for a best-fit slope of 10 degrees. The noise-free case in Figures \ref{fig:roughness_constraints}a and b clearly shows difference in MSR among various roughness. For instance, 20\% of footprints with roughness of 0.9–1.1 m have MSRs exceeding 1 mV$^2$, whereas MSR is always lower than 1 mV$^2$ at roughness below 0.1 m. Therefore, when the MSR larger than 1 mV$^2$ is obtained during a real observation, roughness below 0.1 m can be ruled out. In addition, taking statistics of MSRs within a certain region, comparison in the cumulative probability among roughness bins may be used to indicate the most-likely roughness in the region.\\

Even when realistic noise levels are included, this approach can still constrain surface roughness at the footprint scale. Figures \ref{fig:roughness_constraints}c–f show the same analyses performed under the dayside and nightside noise conditions. Because noise obscures the differences between the best-fit pulse and CW samples, the cumulative probability becomes less sensitive to surface roughness, making roughness detection based on MSR more challenging. Nevertheless, under nightside conditions, the cumulative probability exhibits a clear distinction between roughness below 0.1 m and above 1 m. Therefore, by stacking observations acquired at similar altitudes and slope angles and statistically comparing them with simulated results, CW samples can be used to infer relative roughness variations across different regions of Mercury. For example, rocky terrains in the Apollo-17 DTM exhibit roughness exceeding 0.5 m when they are seen from altitudes of 500 km. Even a lower-bound constraint on roughness may help distinguish rocky areas from areas covered only by regolith on Mercury.\\

It should be noted that the applicability of this method is limited to relatively low altitudes and high surface reflectance, because the MSR is strongly affected by background noise. In contrast, larger surface slopes extend the applicable altitude range, as pulses from steeper slopes are broader and contain more samples that deviate from the best-fit pulse. By examining the slope of the 99\% contour of the cumulative probability maps as a function of roughness at different altitudes, we find that the characterization method based on MSR remains effective up to an altitude of approximately 700 km for a nominal reflectance of 0.17. Although this approach is not applicable to surfaces covered by dark deposits even at the MPO periapsis, it remains effective for icy PSRs, where roughness constraints are achievable even at altitudes exceeding 1000 km.

\section{Implications to laser altimetry on the BepiColombo and other missions}
\subsection{Discussion on the BELA observation and operation}
Our performance simulation provides updated range-error estimates from previous studies by \citet{steinbrugge_performance_2018} and \citet{hosseiniarani_comprehensive_2021}. The range errors reported in this study are more realistic than previous estimates because of improvements in the model assumptions. First, the multiplication gain adopted in this simulation is 20\% lower than those used in previous studies, reflecting the current BELA operation plan. Second, we incorporate laboratory-measured temporal pulse shapes and spatial beam profiles of the transmitted laser pulse for the first time. Third, the realistic modification of the return pulse shape by within-footprint topography is implemented to the prediction in this study. Fourth, range errors due to pointing deviations are refined by explicitly accounting for the actual spatial beam profile. Consequently, the range-error budget reported in this study provides a more reliable basis for future method developments in topographic model construction and tidal deformation measurements as shown by \citet{desprats_mla_2026}.\\

Our performance simulation also demonstrates that BELA can measure return pulse energy with sufficient accuracy for geologic characterization of Mercury’s surface. One of the most important targets of the BepiColombo mission is PSRs, which are generally inaccessible to passive optical instruments such as onboard cameras. Given that the shot-to-shot variation of the transmitted pulse energy is less than 1.4\% \citep{althaus_bela_2019}, the expected uncertainty in relative reflectance is small enough to distinguish dark deposits, regolith, and ice at altitudes below 900 km. Moreover, the mixing ratio of ice and regolith within each footprint may be detectable with an error of 10\% or less during low-altitude MPO operations. Laboratory measurements of the reflectance of ice–regolith mixtures by \citet{yoldi_visnir_2015} show that an ice concentration of 75\% increases reflectance by only 15\%. Such subtle variations are nevertheless measurable at altitudes of 700 km or lower, potentially enabling detailed mapping of ice distribution in Mercury’s polar regions, including micro cold traps \citep{rubanenko_ice_2018}. In addition, the degree of space weathering may be characterized using BELA observations, as demonstrated previously with MLA data \citep{deutsch_temperature-related_2024}. Because the reflectance difference between fresh crater ejecta and surrounding mature material can reach 30\% \citep{riner_spectral_2012}, BELA reflectance measurements can help constrain surface weathering as a clue to relative geologic age of the surface, combined with roughness measurements.\\

It should be noted that the uncertainties in return pulse energy estimated here correspond to relative reflectance, not absolute reflectance, because the transmitted pulse energy itself needs to be calibrated during BELA operations. Throughout the entire BELA operations, degradation of the transmitted pulse energy is expected. According to the BELA life model presented by \citet{kallenbach_space-qualified_2013}, the transmitted pulse energy may decrease by up to 20\% after several hundred million laser shots, with an additional degradation of ~11.3\% due to gamma-ray and photon irradiation. To convert relative reflectance measurements into absolute values, the temporal evolution of transmitted pulse energy needs to be characterized using cross-over calibration points on Mercury. The number of such cross-over locations is estimated to range from 16 to 60 million, depending on mission duration \citep{steinbrugge_performance_2018}. By comparing return pulse energies from the same cross-over locations at different times, the temporal degradation of BELA’s transmitted pulse energy may be quantified.\\

For the first time in a BELA performance model, our simulation demonstrates that digitized CW samples are essential for constraining within-footprint roughness. As shown in section \ref{subsec:misinterpretation}, non-Gaussian return pulse shapes prevent reliable roughness estimation using pulse width solely. In contrast, CW samples preserve information about pulse asymmetry and skewness, enabling roughness constraints, particularly at altitudes below  700 km. After BepiColombo arrives at Mercury in 2026, the MPO spacecraft will be inserted into an elliptical orbit with a periapsis of 480 km and an apoapsis of 1500 km \citep{benkhoff_bepicolombo_2021}. Over the nominal and extended missions, orbital eccentricity is expected to increase, potentially lowering the periapsis to 200 km \citep{hoschele_orbital_2021}. Therefore, the BELA CW samples, especially during the extended mission, will be critical for quantitative observations of tens-of-meter-scale roughness, which is difficult to be accessed by passive optical imagery quantitatively. In the current operation plan, the CW samples of the first return pulse candidates will be downlinked using a data compression technique developed by \citet{huttig_tabled_2024}, ensuring availability of the analysis method of return pulse shapes demonstrated in this study. \\

Roughness estimates derived from CW samples will provide new insights into Mercury’s geologic evolution once it is combined with other datasets. Surface roughness at baselines over hundreds of meters may be inferred by using height measured at neighboring footprints of BELA measurements as done for MLA data analyses \citep[e.g.,][]{kreslavsky_kilometerscale_2014, fa_topographic_2016, susorney_surface_2017, nishiyama_first_2026}. In addition, surface temperature measurements by the Mercury Radiometer and Thermal Infrared Spectrometer (MERTIS; \citet{hiesinger_studying_2020}) can infer centimeter-scale roughness and rock abundance in comparisons with infrared thermal models as conducted for the Moon \citep[e.g.,][]{bandfield_lunar_2015,davidsson_interpretation_2015,nishiyama_utilization_2022,wohlfarth_advanced_2023}. Roughness estimates based on CW samples can fill the gap between these quantitative morphological observations in the BepiColombo mission, providing a multi-scale view of Mercury’s surface evolution.\\

\subsection{Implication to other laser altimeters}
Our performance model demonstrates that the width of the return laser pulse is not a reliable indicator of footprint-scale surface roughness. This conclusion is consistent with the discrepancy between observed pulse widths and those predicted from DTMs reported by \citet{poole_calibrating_2014} in the context of Martian remote sensing. By comparing pulse widths from the Mars Orbiter Laser Altimeter (MOLA) with high-resolution stereo DTMs derived from HiRISE imagery, they showed that pulse width does not consistently correlate with surface roughness. Our BELA simulations further suggest that this discrepancy arises, at least in part, from non-Gaussian return pulse shapes, which are also expected to affect MOLA data.\\

Although \citet{poole_calibrating_2014} argued that regional roughness can be recovered by downsampling and stacking pulse-width measurements, such approaches require a rigorous translation between stacked pulse width and surface roughness. As demonstrated by our analysis of pulse width compared to local slope (section \ref{subsec:misinterpretation}), the stacked pulse width does not agree with pulse widths predicted from DTM-derived surface roughness. Therefore, while this stacking approach has been widely applied to MOLA \citep{neumann_mars_2003} and Lunar Orbiter Laser Altimeter (LOLA; \citealt{smith_initial_2010}) datasets, the interpretation of stacked pulse widths as a direct measure of surface roughness should be revisited. Our analysis suggests that existing pulse-width-based roughness estimates need to be reinterpreted and corrected using realistic pulse-shape simulations, such as those presented in this study.\\

Our pulse-shape simulation framework is directly applicable to future laser altimetry missions that acquire return pulse shapes. For example, the Ganymede Laser Altimeter (GALA) onboard ESA’s Jupiter Icy Moons Explorer (JUICE) mission measures return pulse shapes in a manner fundamentally similar to BELA \citep{hussmann_ganymede_2025,enya_ganymede_2022}. One of the primary science objectives of GALA is to characterize local surface roughness within 50-m footprints \citep{kimura_science_2019}, which may reflect local geologic processes such as cryovolcanism. To date, GALA has measured only dark noises without return pulse data to calibrate its boresight using solar noise data \citep{nishiyama_-flight_2026}. However, during future flybys of icy moons in the Jovian system and after the orbit insertion at Ganymede, GALA is expected to acquire return pulse data with improved conditions for pulse-shape characterization. Because of GALA’s higher ADC sampling rate than BELA and the higher surface reflectance of Ganymede than Mercury, return pulse shapes are expected to be more resolved with a higher signal-to-noise ratio. Therefore, the pulse-shape-based surface characterization methods developed in this study may be adopted for icy moons to study their geologic evolution in the future.\\

\section{Conclusion}
\label{sec:conclusion}
In this study, we developed the most realistic performance simulation of BELA to date by incorporating return pulse modification by footprint-scale topography and the analog filter in the receiver system, combined with laboratory-measured temporal pulse shapes and spatial beam profiles. This comprehensive BELA performance model provides updated and realistic estimates of range errors from multiple sources, including a refined calibration of range bias introduced by analog filtering in the receiver chain. Our results demonstrate that return pulse energy can be measured with sufficient accuracy for geologic characterization of Mercury’s surface, particularly within PSRs, enabling future constraints on the spatial distribution of volatiles such as water ice and organic materials.\\

In contrast to the range and reflectance measurements, our pulse shape simulations reveal that within-footprint roughness cannot be reliably inferred from pulse width data solely. This limitation arises from the fact that return pulse shapes are not simply broadened by surface roughness, contrary to assumptions commonly adopted in previous theoretical frameworks. We further show that straightforward averaging of pulse width data does not infer true surface roughness. As a result, pulse width measurements acquired by previous lunar and planetary laser altimetry missions should be reinterpreted using realistic pulse shape models specific to each instrument.\\

Instead of relying on pulse width data, digitized pulse shape data provide new opportunities for estimating surface roughness at footprint-scale baselines. By comparing CW samples with a library of pulse shapes simulated with idealized flat slopes, a lower bound on within-footprint roughness may be placed. Once BELA obtains real return pulse data, further comparison with other pulse shape libraries, such as return pulses simulated with various fractal DTMs, may pose additional constraints on within-footprint roughness and characterization of Mercury's topographies. The future implementation of pulse shape digitization, as developed for BELA, GALA, and GLAS, will advance the capabilities of laser altimeters to measure footprint-scale roughness, providing insightful information to surface processes and geologic activity across a wide range of planetary bodies.

\section*{List of abbreviations}
\label{sec: abbreviations}
ADC: Analog-to-Digital Converter\\
AEU: Analogue Electronics Unit\\
APD: Avalanche Photodiode\\
APD-A: Avalanche Photodiode Assembly
BELA: BepiColombo Laser Altimeter\\
CW: Correlation Window\\
DTM: Digital Terrain Model\\
EMC: Electromagnetic Compatibility\\
ESA: European Space Agency (ESA)\\
FOV: Field of View\\
FWHM: Full Width at Half Maximum\\
GALA: Ganymede Laser Altimeter\\
GLAS: Geoscience Laser Altimeter System\\
GRM: Ground Reference Model\\
JAXA: Japan Aerospace Exploration Agency\\
JUICE: Jupiter Icy Moons Explorer\\
LHB: Laser Head Box\\
LOLA: Lunar Orbiter Laser Altimeter\\
LROC: Lunar Reconnaissance Orbiter Camera\\
MLA: Mercury Laser Altimeter\\
MDIS: Mercury Dual Imaging System \\
MERTIS: Mercury Radiometer and Thermal Infrared Spectrometer\\
MMO: Mercury Magnetospheric Orbiter\\
MOLA: Mars Orbiter Laser Altimeter\\
MPO: Mercury Planetary Orbiter\\
MSR: Mean Squared Residual\\
PFD: Probability of False Detection\\
PSR: Permanently Shadowed Region\\
RFM: Range Finder Module\\
RMS: Root-Mean-Square\\
SNR: Signal-to-Noise Ratio\\
TIA: transimpedance amplifier\\

\section*{Declarations}

\section*{Availability of data and materials}

All simulation data produced during this study are available from the corresponding author upon reasonable request.

\section*{Competing interests}
The authors declare that they have no competing interests.

\section*{Funding}
This work is supported by the IGPEES WINGS Program of the University of Tokyo, JSPS KAKENHI Grant Number JP22J12387, JP22K21344, and JP26KJ0001, JSPS Overseas Challenge Program for Young Researchers, and JSPS Overseas Research Fellowship. We acknowledge support from the ``Institut National des Sciences de l'Univers'' (INSU), the ``Centre National de la Recherche Scientifique'' (CNRS) and ``Centre National d'Etudes Spatiales'' (CNES) through the ``Programme National de Plan{\'e}tologie''. This work was supported by the Île-de-France Region with DIM origines. This work has received support from France 2030 through the project named Académie Spatiale d'Île-de-France (\url{https://academiespatiale.fr/}) managed by the National Research Agency under bearing the reference ANR-23-CMAS-0041. Open Access funding is enabled and organized by Projekt DEAL.

\section*{Authors' contributions}
GN, AS, CH, and HH conceptualized this study and developed codes for numerical simulations. GN, AS, CH, KW, CA, and TB conducted laboratory measurements. GN conducted all the numerical simulations, wrote the original draft, and finalized the manuscript. All the authors reviewed and approved the manuscript.

\section*{Authors' information}
Nothing to declare.

\acknowledgments{This work is supported by the IGPEES WINGS Program of the University of Tokyo, JSPS KAKENHI Grant Number JP22J12387, JP22K21344, and JP26KJ0001, JSPS Overseas Challenge Program for Young Researchers, and JSPS Overseas Research Fellowship. We thank the BELA Experiment teams at DLR (Institute of Space Research, Berlin) and at University of Bern (Physikalisches Institut, Bern) as well as the BELA Science Team. We also acknowledge the support by the BepiColombo project teams at ESTEC, ESOC, and ESAC.
We acknowledge support from the ``Institut National des Sciences de l'Univers'' (INSU), the ``Centre National de la Recherche Scientifique'' (CNRS) and ``Centre National d'Etudes Spatiales'' (CNES) through the ``Programme National de Plan{\'e}tologie''. This work was supported by the Île-de-France Region with DIM origines. This work has received support from France 2030 through the project named Académie Spatiale d'Île-de-France (\url{https://academiespatiale.fr/}) managed by the National Research Agency under bearing the reference ANR-23-CMAS-0041.  Open Access funding is enabled and organized by Projekt DEAL.

}

\appendix{Temperature effect on return pulse shape}
\label{sec:pulse_temperature}
The temporal shape and spatial profile of transmitted laser pulses may vary due to thermal distortion of BELA. Laboratory measurements by \citet{althaus_bela_2019} observed multiple patterns of $S_T(t)$ and $D_T(x,y)$ under different temperature conditions. In this study, these measurements of pulse shapes were incorporated into the performance simulation to assess thermal effects. Measuring the temperature of the Laser Head Box (LHB), which houses the laser diodes, \citet{althaus_bela_2019} characterized $S_T(t)$ and $D_T(x,y)$ for three temperature conditions: hot, ambient, and cold (see definition by \citet{althaus_bela_2019}).\hspace{0pt}\newline

\begin{figure}
\centering
    \includegraphics[width=0.8\linewidth]{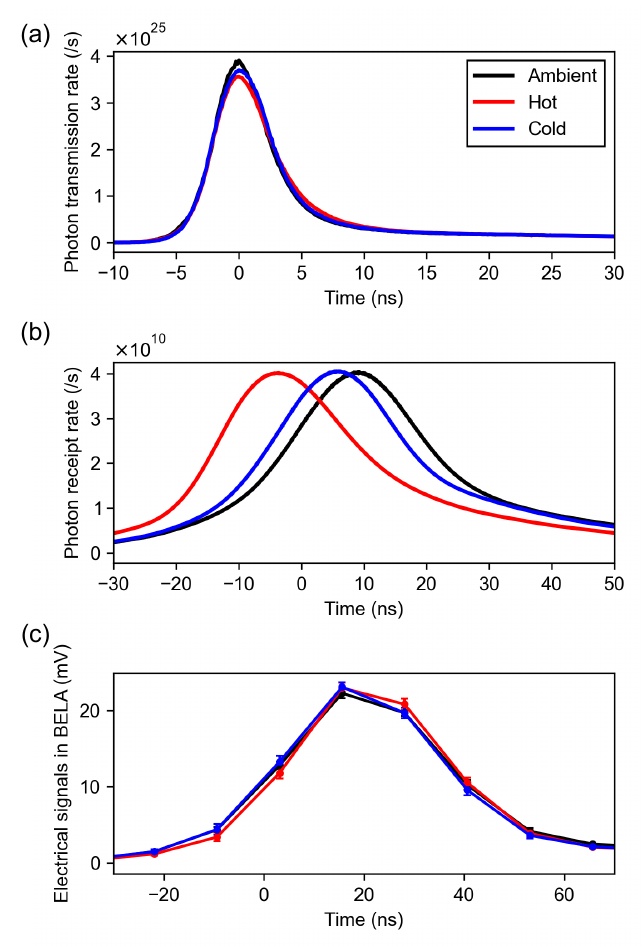}
  \caption{Pulse shape comparison under different temperature conditions. (a) Photon transmission rate from BELA. The total number of transmitted photons is identical for all temperature cases. (b) Photon reception rate at the BELA telescope. (c) Return pulse signal digitized by the BELA ADC. Pulse peak positions are aligned to facilitate comparison of pulse shapes.}
  \label{fig:temperature_pulse_shapes}
\end{figure}

Incorporating laboratory-measured  $S_T(t)$ and $D_T(x,y)$, our simulations reveal that the temperature conditions have little influence on the return pulse shape. Figures \ref{fig:temperature_pulse_shapes}a–c show the transmitted, received, and digitized pulse shapes for the three LHB temperature conditions. The total number of transmitted photons is identical for all cases (Figure \ref{fig:temperature_pulse_shapes}a). The laser pulse is assumed to be reflected from a flat surface with a slope of 10°, corresponding to the expected average slope at footprint scales \citep{hosseiniarani_comprehensive_2021}. Because temperature-induced pointing deviations shift the laser transmission direction, the peak position of the return pulse depends on the LHB temperature (Figure~\ref{fig:temperature_pulse_shapes}b). This variation may introduce bias in range measurements if no correction of temperature conditions is applied. However, its impact on the digitized pulse shape is limited after ADC sampling at 12.5 ns intervals. In Figure \ref{fig:temperature_pulse_shapes}c, return pulses are aligned by their peak positions to facilitate comparison only in pulse shape. The temperature-induced variation in pulse shape reaches up to 2 mV, which is comparable only to the noise level. Because this variation slightly exceeds the noise, temperature effects should be considered when comparing BELA observations with a pulse list (section \ref{subsec:topography effect}). Nevertheless, the effect on reflectance measurements is minimal, as the peak amplitudes are nearly identical among the three temperature conditions.\hspace{0pt}\newline

\begin{figure}
\centering
    \includegraphics[width=1\linewidth]{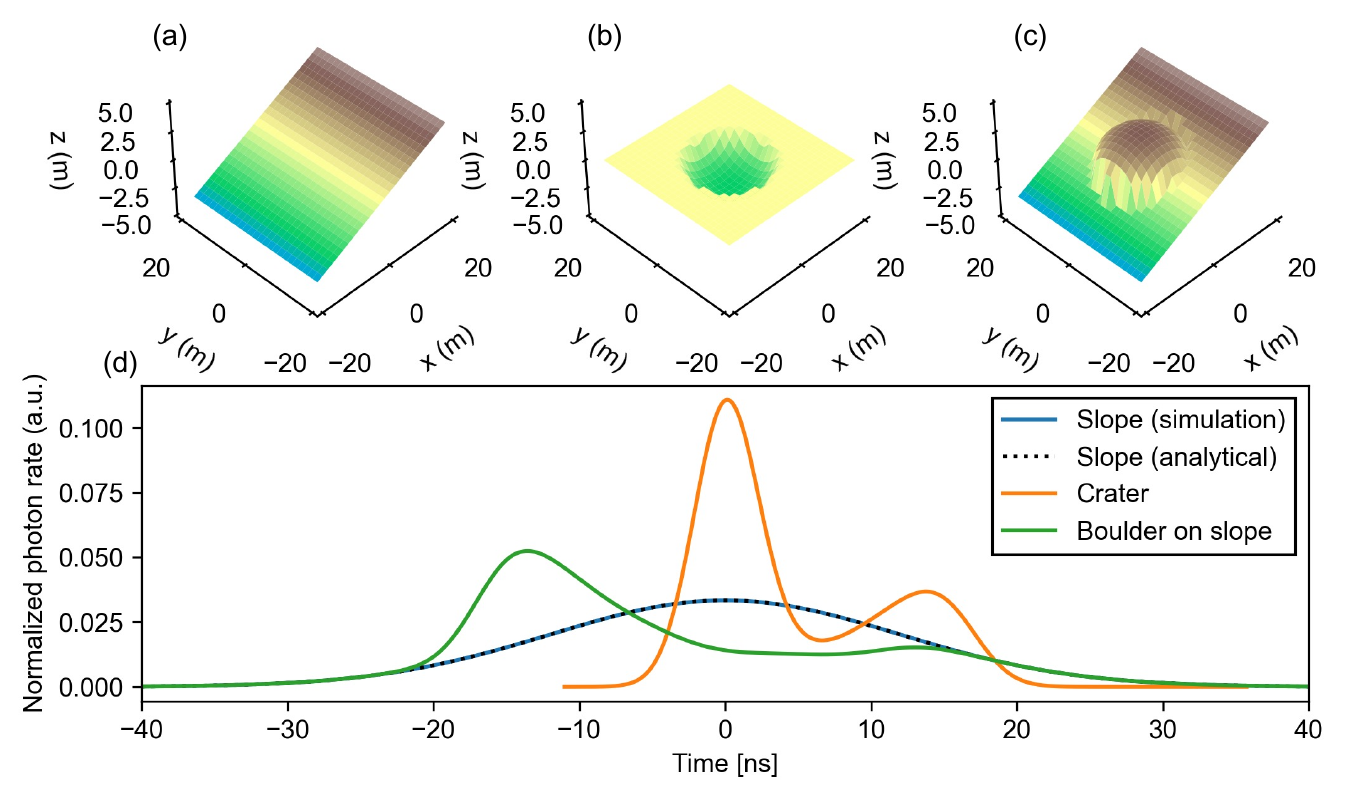}
  \caption{Pulse shape examples. (a-c) Idealized topography models used to validate our performance simulation. (a) a flat slope, (b) a crater, and (c) a boulder on a slope. The slope angle in (a) and (c) is set to 10 degrees. (d) Simulated return pulse shapes for each case. The return pulse from the flat slope (a) is compared with the analytical formula (Eq. \ref{eq:slope_broadening}). Note that the pulse shapes are normalized by the total number of photons of the return pulses.}
  \label{fig:pulse_shape_examples}
\end{figure}

\appendix{Validation of pulse shape simulation with analytical solutions}
\label{sec:pulseshape_detail}
The numerical performance simulation code was validated using an analytical solution for a simple case. As described by \citet{gardner_ranging_1992}, the analytical return pulse shape from a tilted flat surface can be derived by assuming Gaussian temporal pulse shapes and spatial beam profiles:
\begin{linenomath}
\begin{align}
S_T(t) &= \frac{1}{\sqrt{2\pi}\sigma_{trans}}
\exp\left(-\frac{t^2}{2\sigma_{trans}^2}\right), \\
D_T(x,y) &= \frac{1}{2\pi\sigma_r^2}
\exp\left(-\frac{x^2 + y^2}{2\sigma_r^2}\right),
\end{align}
\end{linenomath}
where $\sigma_r$ is the standard deviation of the spatial photon distribution (i.e., $\sigma_r = H \tan \theta_T$). The temporal and spatial coordinate origins are defined at the centers of the Gaussian distributions, and $D_T(x,y)$ is assumed to be axisymmetric for simplicity.
For a flat surface with a radiation factor following the Lommel–Seeliger law, incorporation of Eqs. \ref{eq:photon_basic}, \ref{eq:time_delay}, and \ref{eq:received_apd} yields the photon rate received by the APD as
\begin{linenomath}
\begin{align}
\label{eq:slope_broadening}
n_R^{APD}(t) &=
\frac{\alpha_N \Omega_R \epsilon_0 \epsilon_t N_0}{\pi}
\int \int S_T\left(t - \frac{2x \tan S}{c}\right)
D_T(x,y)  dx  dy \\
&= \frac{\alpha_N \Omega_R \epsilon_0 \epsilon_t N_0}
{\pi\sqrt{2\pi}\sigma_{sum}}
\exp\left(-\frac{t^2}{2\sigma_{sum}^2}\right),
\end{align}
\end{linenomath}
where $\sigma_{sum} = \sqrt{\sigma_{trans}^2 + \sigma_r^2 (4/c^2)\tan^2 S}$, and $S$ is the surface slope angle. Here, the surface is assumed to be inclined only in the $x$-direction. As a benchmark of the code, this analytical solution was compared with numerical simulations for a flat slope (Figures \ref{fig:pulse_shape_examples}a and d). The two results are identical, validating the numerical simulation.\hspace{0pt}\newline

Our simulation further demonstrates that return pulse shape data are essential for investigating within-footprint roughness, as complex pulse shapes can arise even from simple combinations of surface topographies. Figure \ref{fig:pulse_shape_examples} shows three examples of return pulse shapes derived from idealized surface models. When a footprint contains a crater-like depression (Figure \ref{fig:pulse_shape_examples}b), a double-peaked return pulse is generated due to delayed returns from the crater floor (orange line in Figure~\ref{fig:pulse_shape_examples}d). A boulder-like feature produces a pulse shape that is horizontally inverted relative to the crater case (green line in Figure~\ref{fig:pulse_shape_examples}d). Although the average slope angle is identical for the flat-slope and boulder cases (Figures \ref{fig:pulse_shape_examples}a and c), the increase in within-footprint roughness does not simply broaden the return pulse. Instead, it can skew the pulse shape significantly and rather make the apparent shape narrower by focusing pulse energy in one side. Under nominal BELA operation, digital matching with a Gaussian template may therefore measure the width of the narrower peak produced by the boulder, underestimating pulse widths compared to flat slope cases (Figure \ref{fig:pulse_shape_roughness}c).\\

\appendix{Analog filter characterization based on GRM}
\label{sec:butterworth}
To characterize the analog filter of the BELA AEU, we use the BELA GRM located at DLR Berlin as a representative of the BELA instrument aboard MPO. Because the GRM is an assembly of flight spares of electronic units, the functionality of the AEU of the GRM is expected to be identical to that of the AEU aboard MPO. In this laboratory test campaign, an external function generator was connected to the AEU of the GRM, and full-range window samples were acquired using the entire receiver chain. By inserting sinusoidal signals from the function generator and sweeping the signal frequency, the ratio between the input and output signal amplitudes was used to characterize the gain factor of the AEU and its frequency dependence. To investigate a possible dependence on input amplitude, measurements were conducted with peak-to-peak amplitudes of 3, 13, 30, 50, and 200 mV.\hspace{0pt}\newline

\begin{figure}[ht]
\centering
    \includegraphics[width=1\linewidth]{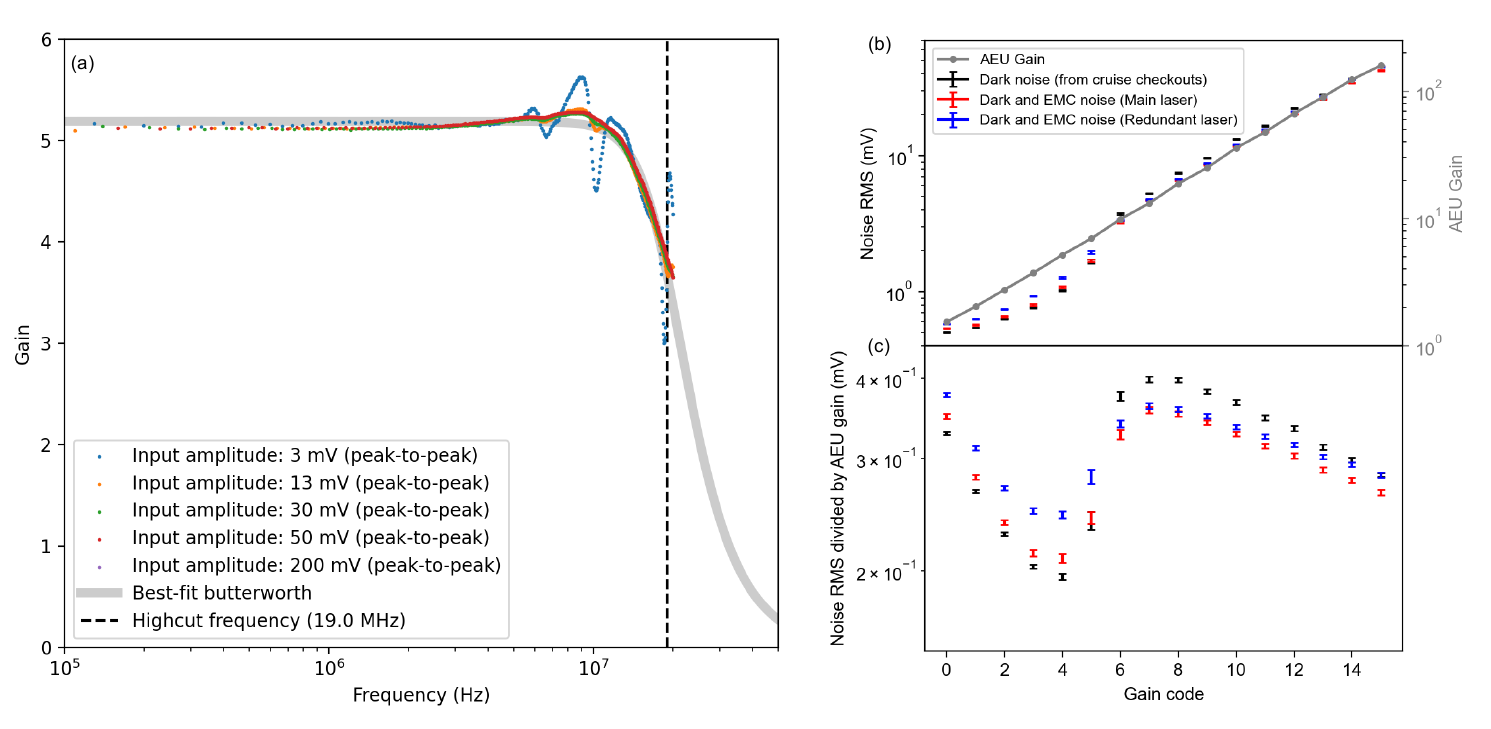}
  \caption{Characteristics of AEU gain. (a) AEU gain amplification as a function of input signal frequency at gain code 4. The color points show each measurements with GRM. The gray shades shows the best-fit Butterworth filter of AEU used in the simulation model. (b) AEU gain compared with dark and EMC noise level as a function of gain codes. (c) Noise RMS divided by AEU gain over the entire range of gain codes. The minimum noise level normalized by AEU gain corresponds to the highest signal-to-noise ratio.}
  \label{fig:butterworth_filter}
\end{figure}

Figure~\ref{fig:butterworth_filter}a shows the measured gain at gain code 4. The gain values are primarily dependent on input signal frequency, decreasing at frequencies above 10 MHz. The gain also varies with input signal amplitude, although the variation is consistent for input amplitudes above 13 mV. By fitting the frequency response with a third-order Butterworth filter, the high-cut frequency was estimated to be 19.0 MHz. Using the gain plateau at frequencies below 10 MHz, the gain amplification factor was estimated to be 5.19. Although our performance simulation considers only gain code 4, the same procedure was applied to all gain codes from 0 to 15 to estimate gain amplification factors (Figure \ref{fig:butterworth_filter}b).\hspace{0pt}\newline

The estimated gain amplification was then used to identify the optimal gain code in terms of signal-to-noise ratio. The dark noise of BELA was characterized in pre-launch laboratory measurements and during cruise checkouts after launch. Because laser transmission introduces additional EMC noise, the dark-noise level measured in the laboratory is higher than that measured during cruise. By dividing the combined dark and EMC noise levels by the AEU gain for each gain code, the noise RMS normalized by the gain amplification was calculated (Figure \ref{fig:butterworth_filter}c). Based on these signal-to-noise ratio estimates, gain code 4 was selected as the nominal setting for future observations and was used throughout this study.\hspace{0pt}\newline

\appendix{Time delay of transmitted pulse}
\label{sec:tx_delay}
The time delay of the transmitted pulse is a key parameter to calibrate the range bias introduced by the analog filter in the BELA receiver chain. The time delay occurs in the same way as that of the received pulse (see Section \ref{sec:arrival_time_bias}). However, the time delay of the transmitted pulse needs to be treated differently from the received pulse because of a discrepancy in the measured and simply-simulated shape of the transmitted pulse detected by the RFM.\hspace{0pt}\newline

\begin{figure}[ht]
  \centering
    \includegraphics[width=1\linewidth]{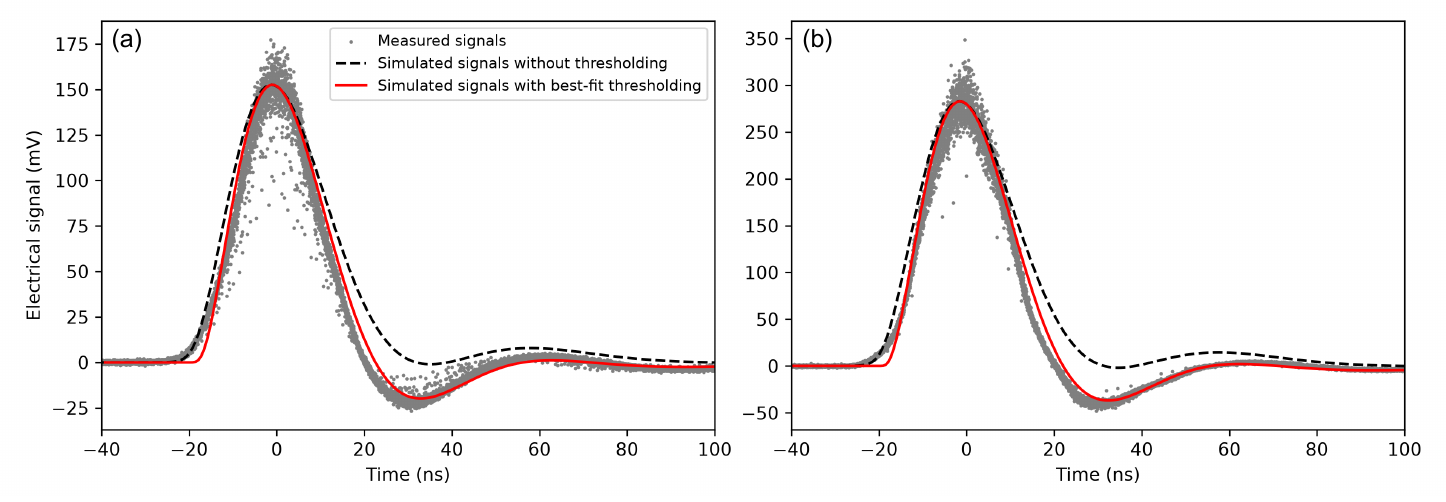}
    \caption{Comparison in electrical signals between laboratory-measured data and simulations using our pulse shape model. The gray points show the measured data after auto-correlation processing. The black dashed and red solid lines show simulated signals with and without thresholding the optical shape of the transmitted pulse. (a) Results for the main laser. (b) Results for the redundant laser.}
    \label{fig:tx_delay}
  \end{figure}

Pre-launch laboratory measurements of the transmitted pulse revealed a discrepancy between the measured and simulated shape of the transmitted pulse detected by the RFM. Figure \ref{fig:tx_delay} shows the comparison in electrical signals between laboratory-measured data and simulations using the analog filter model (Appendix \ref{sec:butterworth}). Although the electrical signals are sampled only every 12.5 ns, auto-correlation among all RFM-detected samples enables the extraction of the electrical signals of the transmitted pulse at higher temporal resolution (gray points in Figure \ref{fig:tx_delay}). Due to the high order of the analog filter (Appendix \ref{sec:butterworth}), the measured electrical signal shows a negative tail next to the peak of the transmitted pulse. However, the simulated signal using the same shape as the laboratory measurement by \citet{althaus_bela_2019} does not reproduce the measured shape of the transmitted pulse well (black dashed line in Figure \ref{fig:tx_delay}), particularly at the negative tail.\hspace{0pt}\newline

This discrepancy suggests that the optical shape of the transmitted pulse is not kept as measured in the laboratory when the pulse is transferred from the transmitter to the APD through the optical fiber. The outgoing shape of the optical transmitted pulse is not a narrow Gaussian but has a long tail due to the decay time of laser pulses (Figure \ref{fig:schematic_method}b), which prevents from creation of the negative tail seen in the RFM-detected samples. Therefore, for consistency between the optical pulse shape and electrical samples detected by RFM, the optical transmitted pulse transferred from the LHB to the APD through the optical fiber would not have the long tail after the peak. The loss of pulse energy in the optical fiber is characterized by a fractional loss dependent on the distance but could pose a threshold of the laser pulse energy that may be transferred. The former process does not change the pulse shape in time and is not consistent with the measured electrical signals. Therefore, in this study, the latter case is considered to make the model consistent with the RFM-measured samples.\hspace{0pt}\newline

Introducing the threshold level to the optical shape of the transmitted pulse, the RFM-measured samples are reproduced (red solid line in Figure \ref{fig:tx_delay}). Changing the threshold from 0 to 100\% of the maximum value of the optical pulse, the optical pulse shape is first modified to have 0 values below the threshold. Then, the analog filtering is applied to the optical pulse shape and compared with the RFM-detected samples to find the best-fit model. Repeating the procedure for the main and redundant lasers, we found the best-fit threshold levels of 17\% and 22\%, respectively. Using this model, the time delay of the transmitted pulse due to the analog filtering is estimated to be 5.49 and 5.45 ns for the main and redundant lasers, respectively (See Section \ref{sec:arrival_time_bias}).

\section*{Endnotes}

\textbf{Figures}

\bibliographystyle{plainnat}
\bibliography{Mercury}
\end{document}